# The Hierarchical Morphotope Classification
## A Theory-Driven Framework for Large-Scale Analysis of Built Form

Martin Fleischmann[1]*, Krasen Samardzhiev[1], Anna Brázdová[1], Daniela Dančejová[1] & Lisa Winkler[2]

[1]Charles University, Department of Social Geography and Regional Development
[2]University of Freiburg, Chair of Environmental Meteorology, Faculty of Environment and Natural Resources

Built environment, formed of a plethora of patterns of building, streets, and plots, has a profound impact on how cities are perceived and function. While various methods exist to classify urban patterns, they often lack a strong theoretical foundation, are not scalable beyond a local level, or sacrifice detail for broader application. This paper introduces the Hierarchical Morphotope Classification (HiMoC), a novel, theory-driven, and computationally scalable method of classification of built form. HiMoC operationalises the idea of a morphotope – the smallest locality with a distinctive character – using a bespoke regionalisation method $SA^3$ (Spatial Agglomerative Adaptive Aggregation), to delineate contiguous, morphologically distinct localities. These are further organised into a hierarchical taxonomic tree reflecting their dissimilarity based on morphometric profile derived from buildings and streets retrieved from open data, allowing flexible, interpretable classification of built fabric, that can be applied beyond a scale of a single country. The method is tested on a subset of countries of Central Europe, grouping over 90 million building footprints into over 500,000 morphotopes. The method extends the capabilities of available morphometric analyses, while offering a complementary perspective to existing large scale data products, which are focusing primarily on land use or use conceptual definition of urban fabric types. This theory-grounded, reproducible, unsupervised and scalable method facilitates a nuanced understanding of urban structure, with broad applications in urban planning, environmental analysis, and socio-spatial studies.

## 1 Introduction

The structure of the built environment affects how cities, towns and villages are perceived as well as the types of activities they can host. The way buildings are laid out, the patterns of streets, the unique open spaces and transformations we observe when we move between neighbourhoods create experiences allowing us to identify a place (Lynch 1960), navigate through urban spaces (Bongiorno et al. 2021) and influence housing preferences and choices (Hasanzadeh et al. 2019). Urban morphology, a field that primarily focuses on understanding of built up patterns and their evolution (Moudon 1997), has a long history in unpacking the complexity of urban (or general built up, when dealing with rural and peri-urban locations) form. While approaches vary from architectural interpretation (Caniggia and Maffei 2001) to modelling based on remote sensing (Abascal et al. 2022), to overall goal is to identify patterns of development in the built fabric. Attempts to do this range from the detection of local morphological regions (Conzen 1988), prediction of planning archetypes (Chen et al. 2021), or delineation of unique concepts like fringe belts (Whitehand 1967) or informal settlements (Wang et al. 2023). Recently, there has been increased interest in one specific approach – exhaustive classification frameworks.

Researchers use classification methods to reduce the complexity of the built-up environment into types. The resulting typologies in turn are used to better understand the structure of cities and their relationship to other phenomena. For example, the built-up environment's impact on the heat island effect (Wu et al. 2025), energy efficiency (Zhu et al. 2022), broader economic outcomes (Wang et al. 2024), as well as the overall patterns of urbanisation (Domingo et al. 2023; Wang et al. 2024). Furthermore, analysing the interplay of the physical urban fabric with land use (Fleischmann and Arribas-Bel 2022), mobility (Calafiore et al. 2023), vegetation (Guyot et al. 2021) or other dimensions of activities is more manageable when built form can be represented by a small number of interpretable types. However, each of these application domains typically has a different classification method focused on different aspects and scales. Some focus only on characteristics and configurations of buildings (Caruso et al. 2017); or only on streets (Araldi and Fusco 2024); or on total built-up surface within a predefined unit, i.e. patch (Jochem et al. 2020).

These classifications of urban environment have a long history across disciplines and can involve both qualitative and quantitative aspects. Urban planning approaches with a pronounced qualitative component typically provide a detailed account of a small geographical area, based on the particular characteristics of the urban fabric in the area, including social aspects, as well as its history of development (Conzen 1960). Driven by advancements in data processing and computing, recent works have successfully adopted quantitative methods to achieve similar hyper-locality and detail (Fleischmann et al. 2022; Araldi and Fusco 2019, 2024). Such approaches create detailed typologies that differentiate between different historical layers, distinguishing medieval from modernist urban developments, single family housing from row-houses and similar (Arribas-Bel and Fleischmann 2022; Fleischmann et al. 2022).

Nevertheless, a consistent classification that stretches beyond a delimited boundary of a single metropolitan area or a country is still missing. At least three factors have prevented such work being carried out so far – theory, data and methods – each interacting with the others and amplifying the overall difficulty of the task. Theoretical considerations drive applications and both data and methodological choices. However, they are at the same time limited by data availability and methodological scalability. Therefore, theoretically relevant approaches are typically limited by their method to neighbourhoods or multiple cities at most, i.e. (Fleischmann et al. 2022; Dibble et al. 2019). Other purely quantitative approaches generally sacrifice detailed typologies and theoretical justifications, and focus only on a few coarse morphological features in favour of scalability to larger regions (European Environment Agency 2020, 1990; Stewart and Oke 2012).

This paper aims to tackle the three interrelated issues of theory, data and methods by creating a scalable urban morphology classification method that is directly stemming from theoretical foundations of urban planning.

The rest of the paper is structured as follows. The next section positions the overall quantitative urban morphological approach, as well as our specific method, within the literature. The Methods and Data sections describe the approach we have taken in detail, and also provide some background information about the study area. Furthermore, they describe the qualitative and quantitative steps we take to verify the resulting classification – the descriptions of types, the spatial distribution of the results and the quantitative comparisons with other available large scale urban form datasets. We present the outcomes in the Results section, while the Discussion and Conclusions section focuses on the implications of the results, their limitations and the potential for future work.

## 2 Conceptualisation of urban form

Within cities conceptualised as complex adaptive systems (Batty 2012), urban form is a component that typically reflects the configuration of buildings, streets and plot within the built-up landscape (Moudon 1997). These, so called, fundamental elements of urban form have been present in most of the early schools of urban morphology, no matter if they were stemming from architecture (Muratori 1959; Caniggia and Maffei 1979) or geography (Conzen 1960; Whitehand 1967). Researchers have been attempting to group and delineate them to allow discussion about a *type* of urban form and its spatial allocation, since theoretically, there are infinitely many possible configurations. Conzen's *morphological region* (Conzen 1975) or Italian *urban tissue* (Muratori 1959) are among the concepts defining a morphologically homogeneous patch of land, that can be further compared to other such patches, eventually identifying a recurring type. Each uses its own criteria and methods of delineation, but share the same goal. Importantly, Conzen further suggests there is a hierarchy of morphological regions (in a sense of nesting rather than ranking) with the smallest one being called *morphotope* (as in *biotope*), defined as "*[t]he smallest urban localit[y] obtaining distinctive character among their neighbours […]*" (Conzen 1988, 259). This is typically closely linked to the period-determined building types and requires information on the process of historical development within the study area (Arat 2023; Conzen 2018). An analysis with this type of data dependency is therefore cumbersome to scale beyond individual cities as the data capturing development period in which a building has been constructed is scarce and often encoded only in digitised scans of historical maps. At the same time, given the definition of "the smallest" is not easy to define, researchers tend to skip the detection of morphotopes and focus primarily on morphological regions of higher order with a handful of works (e.g. Fedchenko (2023)) aiming to identify morphotopes within urban fabric. Yet, the idea is fundamental and the identification of morphotopes and their hierarchy of morphological regions at different levels of granularity could be used within a wide range of applications, both within and outside urban morphology.

While Conzen's methods are primarily qualitative, based on the interpretation of historical maps and town plans (Oliveira 2019), quantitative approaches are on the rise trying to resolve some of the issues with scalability and reproducibility of morphological analysis. Among such methods, morphometric analysis, stemming from early works of Conzen (1988) or Slater (1981) is gaining more prominence.

Morphometric analysis, which has been originally suggested for numerical analysis of plots based on their detailed measurements (e.g. width in Slater (1981)), has grown into a distinct area of study called *urban morphometrics* (Oliveira and Porta 2025). While the term has been around at least since Porta et al. (2011), it lacks a formal established definition. Here, we offer one and define urban morphometrics as "*a study of urban form through the means of quantitative assessment of its constituent elements*", where constituent elements are typically derived from Moudon's fundamental ones but are further influenced by data availability and the aims of the study. At the same time, the links between morphological theory and quantitative computational methods proposed under the umbrella of urban morphometrics (explicitly stated or not) do not tend to be particularly strong. There are a few exceptions, notably the work of Araldi and Fusco (2019, 2024) stemming from the combination of British and French theories of urban morphology mixed with the streetscape skeleton approach or that of Fleischmann et al. (2022) building on a mixture of British and Italian foundations. We believe that a morphometric assessment needs to be well founded within morphological theory and that such a link should be made explicit, rather than employing the *"data will tell"* approach hoping that the measurement framework is accidentally sensible. Only in this way can we minimise the bias of the morphometric method and make it trustworthy and interpretable.

The methods of urban classification that can be considered morphometric are a relatively recent development, with the earliest attempts dated to late 2000s and early 2010s (Song and Knaap 2007; Colaninno et al. 2011; Gil et al. 2012). While some other related works could be listed here, they usually do not fulfil the definition of urban morphometrics, typically due to reliance on different features than constituent elements of urban form, typically pixels or patches (Huang et al. 2007; Thomas et al. 2010; Wurm et al. 2010), or are very borderline in their methods (Smith and Crooks 2010). Since then, the number of methods have been consistently growing, showcasing the attractiveness of urban morphometrics that is partially driven by the rise in the availability of data capturing morphological elements (e.g. growth of OpenStreetMap (Biljecki et al. 2023)) and partially by the growth of geographic data science (Singleton and Arribas-Bel 2019).

Morphometric classifications started on sub-city scales, focusing on individual neighbourhoods (Gil et al. 2012) or transects through the city (Colaninno et al. 2011). Later, they proceeded to cover entire cities or metropolitan regions (Schirmer and Axhausen 2015; Araldi and Fusco 2019; Bobkova et al. 2019; Jochem et al. 2020; Fleischmann et al. 2022). Only recently, the evolution of methods, availability of computational resources and data resulted in multiple methods that cover entire countries (Araldi and Fusco 2024; Fleischmann and Arribas-Bel 2022). Beyond that scale, classifications of urban form do exist (European Environment Agency 2020; Demuzere et al. 2022) but are typically not of a morphometric origin and a result of remote sensing or other procedures. That pushes them away from morphological theory, yielding classifications that are often too simplistic or archetypal due to a requirement of classes to be defined prior to running the analysis, unlike in a typical morphological assessment where classes stem from observations on the ground.

Methodologically, we can observe rising complexity of morphometric classification methods. First attempts were using a handful of measurements, typically followed by k-Means clustering or similar (Gil et al. 2012; Bobkova et al. 2019), occasionally replaced by more robust clustering algorithms and a higher number of measurements (Fleischmann et al. 2022; Araldi and Fusco 2019). While these approaches remain popular due to their simplicity and good overall performance (Marwal and Silva 2023; Yu et al. 2023; Venerandi et al. 2024), methods derived from neural networks appeared (De Sabbata et al. 2023; Liu and Song 2025; Chen et al. 2025) in an attempt to extract patterns in a more advanced way. Despite these developments, large scale classifications are still hard to come by and those that exist cannot be classified as morphometric.

Typically, global or continental data products make use of Local Climate Zones (LCZ) (Stewart and Oke 2012), which in the case of urban form contain 10 conceptually defined classes. Thanks to the progress in remote sensing, these classes can be predicted across the globe (Demuzere et al. 2022) and then further analysed (Debray et al. 2025) but LCZs are a rough proxy of intricacies of urban form at best, reiterating the point above about the missing link between morphological theory and classification offered.

The current state of morphological and morphometric classifications can henceforth be summarised as follows. The methods that are strongly embedded in theory tend to be not scalable and typically qualitative. The methods that are quantitative (morphometric) and theory-driven (at least partially) have seen an improvements in their ability to scale to national extents but none have been tested on a larger area, raising questions on where their limits are. The methods that show scalability to global extents lack the theoretical foundation in urban morphology and a detail we require from morphological classifications.

The aim of this paper is to develop a method of morphometric classification that is heavily theory-driven, building on Conzen's concepts of morphotope and hierarchical embedding of morphological regions, but completely computational and scalable beyond national boundaries at the same time. To achieve this, we propose a *"Hierarchical Morphotope Classification"* (HiMoC) – a hierarchical method for classifying urban form into a taxonomic tree of algorithmically delineated morphotopes.

## 3 Methods

The HiMoC method can be broken down in eight steps as illustrated in Figure 1. Starting with data retrieval and pre-processing, followed by morphometric characterisation, delineation of morphotopes and their morphometric characterisation, closing with definition of morphological linkage and flexible understanding of urban fabric. Each of the steps is described in detail in the following sections.

### 3.1 Urban form as data

Any morphometric assessment first needs to represent the constituent elements of urban form as data to allow measurement. In our case, we limit the input to two datasets (Figure 1 a). First, building footprints represented as polygons, and second, street networks represented as lines capturing individual road segments. This input captures only a subset of morphological properties (e.g., there is no information on plot boundaries or open space) but is relatively abundant, ensuring that the data availability is not the main limiting factor in applying the method to different contexts. At the same time, this input proves to capture the core essence of morphological patterns (Fleischmann et al. 2022; Araldi and Fusco 2019). The omission of building height data might hinder certain aspects of the classification. However, we argue that we that the absence of height data is not critical since 2-dimensional morphology tends to capture height information (Milojevic-Dupont et al. 2020).

The foundational spatial element on which the classification is done is a building, as the smallest available morphological feature we can work with. From streets and buildings, we further generate enclosed tessellation (Arribas-Bel and Fleischmann 2022), which partitions space into regions delimited by street geometry and distance to buildings. Formally, it is *"the portion of space that results from growing a morphological tessellation within an enclosure delineated by a series of natural or built barriers identified from the literature on urban form, function and perception"* (Arribas-Bel and Fleischmann 2022, 5), where morphological tessellation is a polygon-based Voronoi tessellation with building footprints as an input (Fleischmann et al. 2020). Enclosed tessellation serves as a proxy for the calculation of morphometric attributes on a plot level while, importantly, capturing the open spaces between buildings.

Compared to previous applications of enclosed tessellation, we are not interested in unbuilt open spaces. Therefore we limit the extent of the enclosed tessellation around each building using a variable bandwidth. The bandwidth reflects the assumed size of an underlying plot based on distance to surrounding buildings and is different for each enclosed tessellation cell (see Appendix F).

The street network is split into nodes, representing junctions, and edges, representing segments, as different morphometric attributes are computed for each. The data is further processed to ensure topological correctness of the building layer and simplification of the street network to reflect a morphological, rather than a transportation representation, based on Fleischmann et al. (2025). For the details, refer to Appendix D.

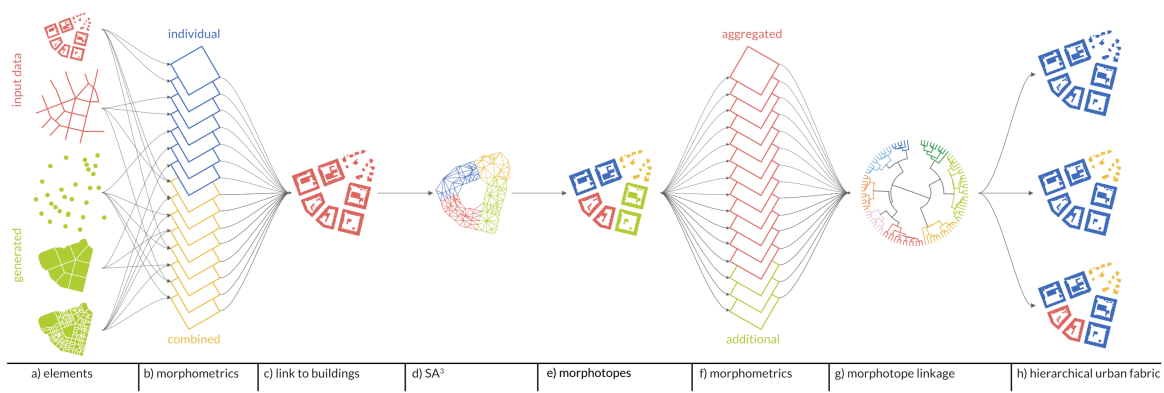

*Figure 1.* Illustration of the methodology

## 3.2 Morphometric characterisation

Measurements are at the core of every morphometric method. While there are potentially hundreds of attributes to measure, not all are equally relevant. Many capture the same concept (e.g. various shape indices reported in Basaraner and Cetinkaya (2017)) or are useful for analysis at scales we are not interested in (e.g. Schirmer and Axhausen's (2015) accessibility metrics). Given that the HiMoC method shares the same input data and general aim as the work of Fleischmann et al. (2022) (apart from building height), we start from their set of "primary" morphometric attributes, remove those depending on height and add a small subset of additional attributes capturing several concepts the original set did not. In total, we measure 59 attributes reported in detail in Appendix A and implemented in a Python package `momepy` (Fleischmann 2019) (Figure 1 b). The set is designed to measure properties of streets, buildings, street intersections and enclosed tessellation cells themselves at three scales: *small*, covering only aspects of the element; *medium*, covering aspects of the element and neighbouring elements; and *large*, covering neighbouring elements up to five topological neighbours; and two subsets: the *individual* layer (e.g. measuring a shared walls ratio of buildings) and the *combination* of layers (e.g. measuring the relation of a building to a street). Each character is measured at the level of individual elements (be it building, street node, or tessellation cell) and subsequently linked to the tessellation layer (Figure 1 c).

## 3.3 Morphotope delineation

Delineating morphotopes based on the morphological attributes of buildings and their surroundings as outlined in Section 3.2 is the first type of aggregation we carry out. Going back to the definition of a morphotope, our goal is to find the smallest urban locality obtaining distinctive character. That implies a strict requirement of contiguity of a morphotope and an application of a clustering method able to define boundaries. To operationalise this step we propose a bespoke regionalisation method called Spatial Agglomerative Adaptive Aggregation ($SA^3$). The algorithm generates a full linkage matrix using Ward's agglomerative clustering restricted to make connections on a contiguity graph of tessellation cells linked to buildings. Then, we extract clusters from the linkage using the density-based cluster extraction method called 'leaf' (McInnes et al. 2017) (Figure 1 d,e). The specific implementation details are provided in Appendix B.

The choice of this algorithm for the delineation of morphotopes is guided by theoretical and computational considerations. First, it produces spatially contiguous clusters. Second, the algorithm admits noise – it can assign a building to a noise cluster if there is not enough of its immediate neighbours in geographical space that are morphologically similar. Third, there can be more morphological heterogeneity within some morphotopes, than between neighbouring morphotopes of different types. For example, a medieval or industrial area has more internal heterogeneity, than two neighbouring morphotopes that represent semi-detached and single family houses respectively. Due to the usage of density-based extraction, $SA^3$ can delineate these types of morphotopes despite their variable internal heterogeneity. In contrast, methods that rely on a flat cut of the full linkage matrix could either split some types of morphotopes if the cut value is too low, or combine them if the value is too high. Fourth, the final number of morphotopes does not need to be specified a priori. The only parameter required by $SA^3$ is the minimum number of buildings to form a morphotope. We observed that very small values (10, 25) tend to produce clusters too granular to be treated as morphotopes, while larger values like 1,000 start to lead to spurious connections. Based on that, we selected a value of 75 to err on the side of avoiding spurious connections, since missing connections can be picked up at later steps. The implementation of $SA^3$ has been contributed to a Python package spopt (Feng et al. 2022) and used from there.

Taken all together, our operationalisation of the concept of a morphotope is: a group of 75 or more directly neighbouring buildings that are more similar to each other morphologically, than to any other such group around them.

### 3.4 Morphotope linkage

Once the morphotopes are delineated, the remaining task is to capture the dissimilarities between between them in order to generate a nested hierarchy. This can be done by using the original morphometric attributes that went into $SA^3$ aggregated per morphotope. We use median value of each, while measuring three additional attributes enabled by the new morphotope aggregation, all outlined in Appendix A (Figure 1 f). We then combine the morphotopes based on similarity only in feature space (no spatial restrictions) to produce a linkage in the form of a taxonomic tree (Figure 1 g). This is done using Ward's agglomerative clustering algorithm. To ensure scalability of the algorithm we generate a 10 nearest neighbours graph (in feature space) for the full dataset and limit the potential Ward distance calculations to this subset. This step only marginally affects the top of the tree while significantly reducing the computational requirements.

The linkage only includes morphotopes and all the buildings labelled as noise in the previous delineation step were excluded. In order to include them in the final analysis, we later assign them to a branch of the taxonomy based on morphometric similarity. To do this we first, group together noise buildings into "noise pseudo-morphotopes" based on spatial contiguity. Then we select a flat cut of the hierarchy, where each branch represents a type with attributes equal to the median value of the morphotope features within it. Lastly, each "noise pseudo-morphotopes" is assigned the most common branch, among its five nearest clusters based on similarity in feature space. Alternatively, all noise points can be discarded.

The final hierarchy shows the dissimilarity between morphotopes, which can be extended all the way down to dissimilarity between pairs of buildings since the $SA^3$ algorithm is also hierarchical. This hierarchy can be cut at arbitrary levels to produce clusters at various scales which represent taxa or urban fabric types, similarly to Conzen's hierarchy of morphological regions.

### 3.5 Interpretation

The hierarchical tree is a powerful representation of morphological similarity, but one that is non-trivial to interpret. With flat clusters as in Fleischmann and Arribas-Bel (2022), one needs to only name and describe each. Within a tree, such an exercise needs to be done with the hierarchy in mind and can quickly get out of hand if we go too deep. To facilitate the interpretation of HiMoC, we propose naming branches up to three levels of bifurcation. For each level, we describe each taxa branch based on the morphological features of the morphotopes within them and provide a short description and pen portrait of each. This naming minimises the complexity of the subsequent analysis and highlights the general patterns in the taxonomic structure.

### 3.6 Evaluation

Evaluating the validity of a classification of this sort is always complicated as there is typically no "ground truth" to compare to. To mitigate this problem to an extent, we use external data capturing adjacent concepts and determine whether the degree of similarity between the our results and the external data is sensible or not. We compare the results of HiMoC at our most granular named level with three existing data products – CORINE Land Cover (European Environment Agency 1990), Copernicus Urban Atlas (European Environment Agency 2020), and Global Map of Local Climate Zones by Demuzere et al. (2022). Each offers a different perspective on our results. CORINE, with its limited granularity of classes that cover the urban environment, provides insight into relation of taxa to urban continuity and specialised land use (industrial zones). Urban Atlas introduces an explicit dimension of density alongside continuity. LCZs theoretically provide the largest granularity of urban classes (10) that take into account compactness, height, density, size and land use but are, in principle, conceptually defined. The underlying assumption is that there should be a clear relationship between the taxonomy and each of these classifications, limited by the different concepts they are built on. To quantitatively analyse the similarity, we create confusion matrices based on spatial intersections, by assigning each building from our hierarchy the appropriate label from each of the three comparison datasets. Our main goal with the comparisons is to demonstrate the usefulness of our results by showing where our clusters overlap and potentially extend existing data products.

Another level of evaluation can be done qualitatively, using visual observation and expert knowledge. This involves a selection of case studies to focus on and understanding whether the patterns captured at various levels represent expert understanding of morphology of the area. Ideally, the selection shall cover different sizes of settlements and multiple geographic regions within the study area. Here, we will present a subset of this exercise focusing on four settlements from population of 8 thousand to 3.4 million. We analyse the patterns of urban form as we move through the three named levels of the taxonomy and the relative importance of different features.

Furthermore, we analyse the spatial distribution of the most granular named level, in order to demonstrate the ability to capture urban form patterns across administrative borders.

It should be noted that we use flat cuts of all subsequent analysis, but multilevel cuts are possible and likely to produce better results for specific applications.

## 4 Data

Our study area covers six contiguous European countries – Austria, Czechia, Germany, Lithuania, Poland, and Slovakia. There are three main reasons for this selection – data availability, morphological diversity, and available expert knowledge.

First, there is official open cadastral data available for every one of these countries as they are all part of the INSPIRE program (Craglia and Annoni 2007) and provide polygons for the entire national building stock in a (relatively) consistent manner. Second, the urban form of Central Europe is highly heterogeneous due to different planning paradigms and approaches applied in history, and due to the presence of well-preserved historical districts as well as those coming from post-war rebuilding. Although not exhaustive, this diversity enables us to evaluate the method on a wide range of urban fabric types. Last, the research team has a large experience of working on morphological research in the region and is able to expertly evaluate the outcomes of the method.

There are two morphological elements required to build the taxonomy – building footprints in the form of two-dimensional polygons and streets in the form of linestrings. The buildings are downloaded from the official cadastres of each country, while the streets are a direct download from Overture Maps. In total, the classification covers over 90 million building footprints and around 21 million street network segments.

## 5 Results

As indicated in the methods, the processing pipeline generates large amounts of data divided into multiple steps. Here we report results from the key parts – morphotope delineation, linkage, interpretation, and evaluation. For detailed insights into morphometric characterisation, please refer to the data product outlined in Section 7.

### 5.1 Morphotope delineation

The $SA^3$ algorithm resulted in the delineation of more than 500,000 morphotopes as spatial aggregation of 90 million buildings. An mean size of a morphotope is 132 buildings, with the largest one covering 636 buildings. An illustrative example covering the city centre of Prague, CZ, is shown in Figure 2.

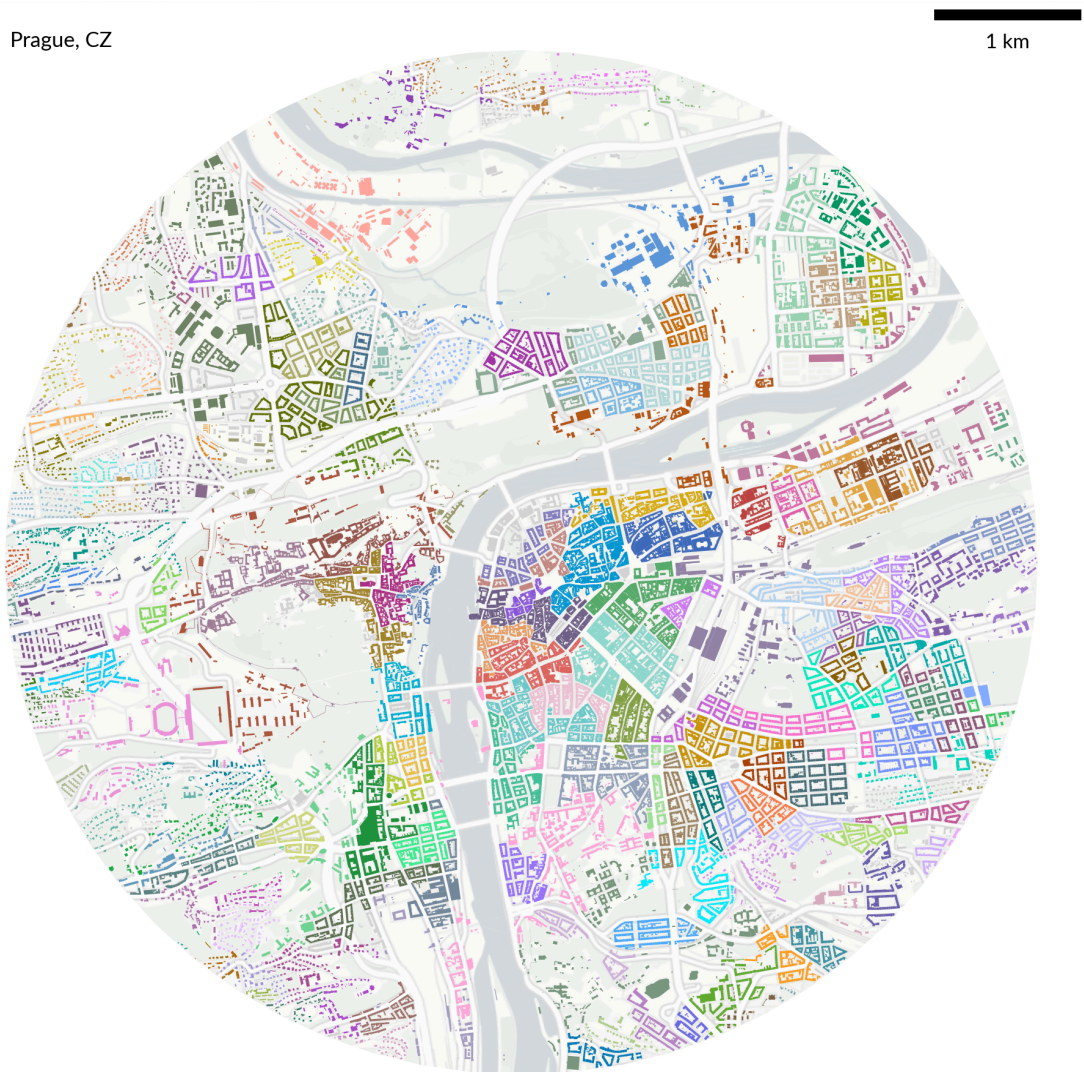

*Figure 2.* Delineation of morphotopes in the city centre of Prague. Each colour captures a single morphotope, while a light grey reflects buildings marked as 'noise' by $SA^3$.

The result tends to capture the concept of morphotope very well from the interpretative perspective - the delineations are typically *the smallest locality obtaining distinctive character* and it is unlikely that any further splitting would be meaningful. What Figure 2 illustrates is the high sensitivity of the algorithm to differences in morphological structure and an excellent ability to follow the edges of distinct patterns, which is something that similar methods typically struggle with (e.g. Fleischmann et al. (2022), Fleischmann and Arribas-Bel (2022), or Araldi and Fusco (2024)). One can argue that some neighbouring morphotopes could be joined. However, having more granular delineations lies closer to the idea of *the smallest* and the subsequent construction of the taxonomic tree will create these further joins.

## 5.2 Morphotope linkage and taxonomy

The linkage between the morphotopes is done using Ward's agglomerative clustering, meaning that the resulting structure is a hierarchical tree capturing dissimilarity between pairs and later groups of morphotopes within its branches. While the full tree has around 500,000 levels, it is typically more practical to look only at the top branches and rely on the bottom ones for specific localised use cases.

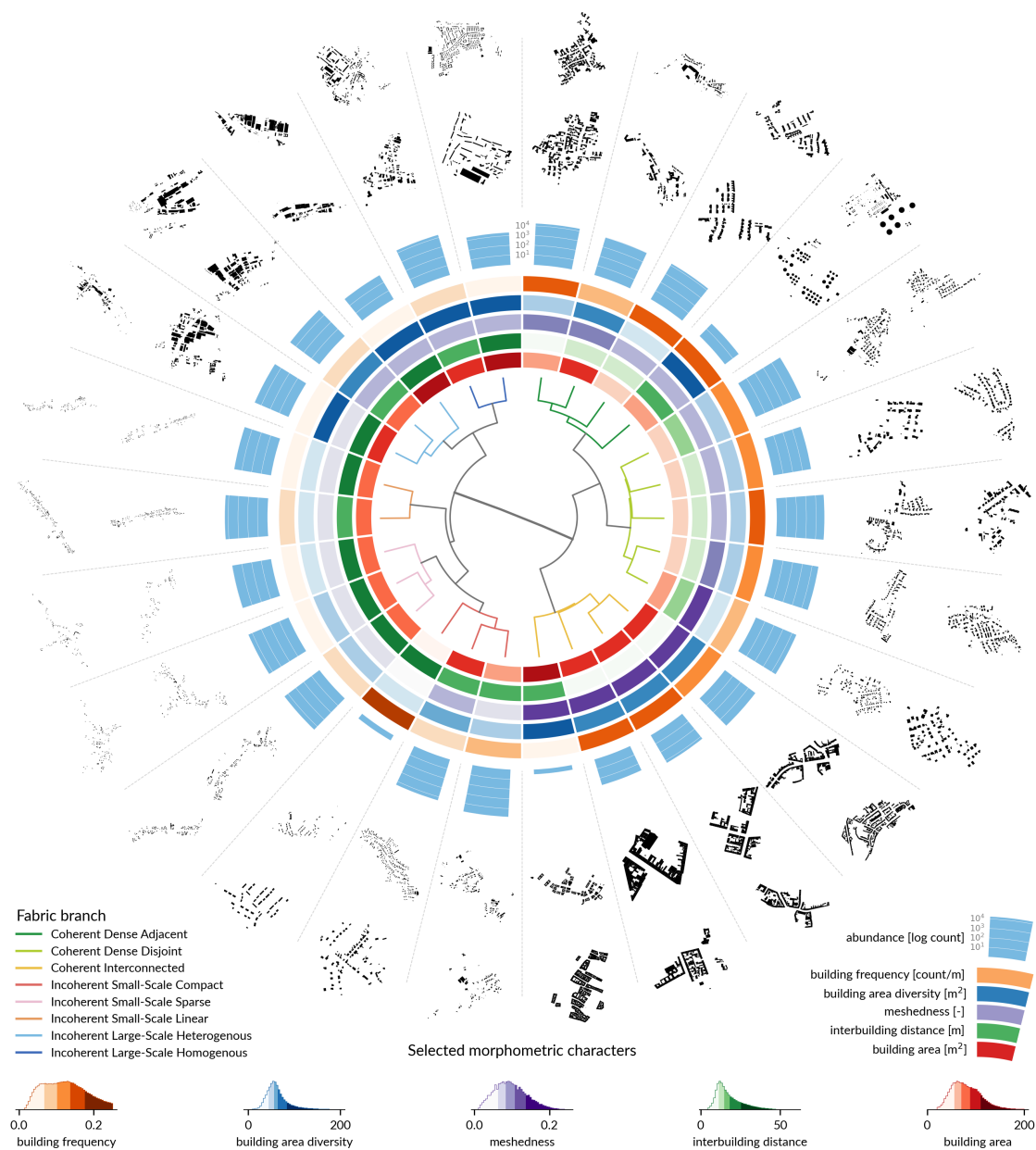

*Figure 3.* Overview of the resulting hierarchical morphotope taxonomy, showing the radial representation of top levels of the taxonomic tree, a sample of their morphometric profile, abundance and two example morphotopes for each branch.

Figure 3 provides a top-level overview of the resulting taxonomy organised around a radial representation of the tree, showing the top 26 branches. The tree is coloured according to the interpretation of the branches at the third bifurcation (see Section 5.3), showing named types of urban fabric. The five rings around the tree provide an insight into a numerical profile of each branch based on a selection of morphometric attributes at the morphotope level. This captures the reasons why the tree looks as it looks and why we consider some morphotopes more similar than others. The legend of attributes further shows the global distribution of values. The next ring represents a bar chart of abundance of each type within the study, represented as the log of the number of morphotopes per branch, highlighting that some types of urban fabric are more prevalent than others and that certain branches can even be considered as extremely rare outliers. The final two rings contain two morphotopes nearest to the centroid of each branch and thus examples of what morphotopes within that branch tend to look like. Note that the morphotopes are not to scale due to their varying spatial extents.

### 5.3 Tree interpretation

We derive names and short pen portraits of each of the eight taxonomic branches until the third level of bifurcation. We do this to improve the subsequent analysis, interpretation and mapping of the different types of urban fabric from the branches. The naming is derived from an expert interpretation of the morphometric profile of individual branches and attempts to stick strictly to this input. While it would be tempting to call certain branches *"medieval city"* or *"industrial area"*, neither historical origin nor land use played any role in the classification and the fact that they are showing in the taxonomy is only a result of their distinct morphological patterns. See Appendix E for the detailed branch profiles, short names and branch descriptions.

The first bifurcation is characterised by the split between what we call *Coherent* and *Incoherent* fabric. This split mirrors the one identified by Dibble et al. (2019) and largely captures the distinction between places where all buildings, streets, plots and blocks have their traditional structural role (e.g. perimeter block of 1800s, see Levy (1999)), and those where at least one of is not significant from a structural perspective (e.g. modernist housing). It reflects the differences between pre- (coherent) and post-war (incoherent) development as well as those between traditional residential areas (coherent) and industrial zones (incoherent). Village development, where a block is only loosely defined, falls typically within the *Incoherent* branch, whereas garden cities fall typically within the *Coherent* branch.

In the second bifurcation, the *Coherent* fabric splits into *Dense* and *Interconnected*, reflecting various degrees of amalgamation, while the *Incoherent* fabric into *Large-scale* and *Small-scale*, capturing the varying scale of constituent elements. The third bifurcation then distinguishes between *Coherent Dense Disjoint* and *Coherent Dense Adjacent* fabrics based on the changes in the degree of building adjacency, between *Incoherent Large-scale Homogeneous* and *Incoherent Large-scale Heterogeneous*, which is driven by internal diversity of buildings and their configurations, and between *Linear*, *Sparse*, and *Compact* branches of *Incoherent Small-scale* fabric, distinguishing different types of typically rural and peri-urban development.

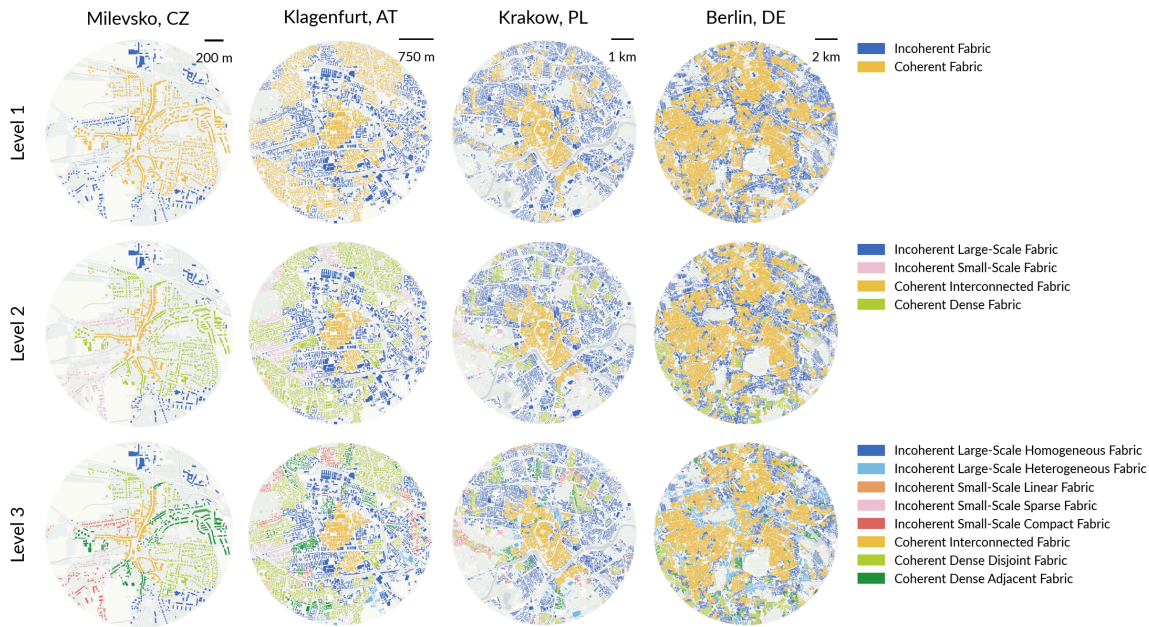

*Figure 4.* Examples of delineation of urban fabric types at different levels of taxonomic granularity and across places of different scale.

Figure 4 shows how these splits manifest in four settlements of different sizes. This example shows that despite significant differences in the nature and geographical location of these places, similar morphotopes are consistently linked together and remain within the same branches of taxonomy.

We can further explore the third bifurcation, with eight named branches in Figure 5, illustrating the outcomes of the classification across major cities in the region. City centres of all major cities fall under *Coherent Interconnected Fabric*, with different extents, from the very small historical core of Bratislava to the large portion of Vienna. The composition of these cities vary due to many reasons. For example, both Prague and Bratislava show high proportions of *Incoherent Large-scale Homogeneous Fabric*, where we typically see large housing developments of the communist period. While some areas falling under the same fabric are present in Munich, and marginally in Vienna, they are much scarcer. Conversely, Vienna, and especially Munich, showcase a much larger proportion of *Coherent Dense Disjoint Fabric*, often covering developments in garden city styles or urban villas.

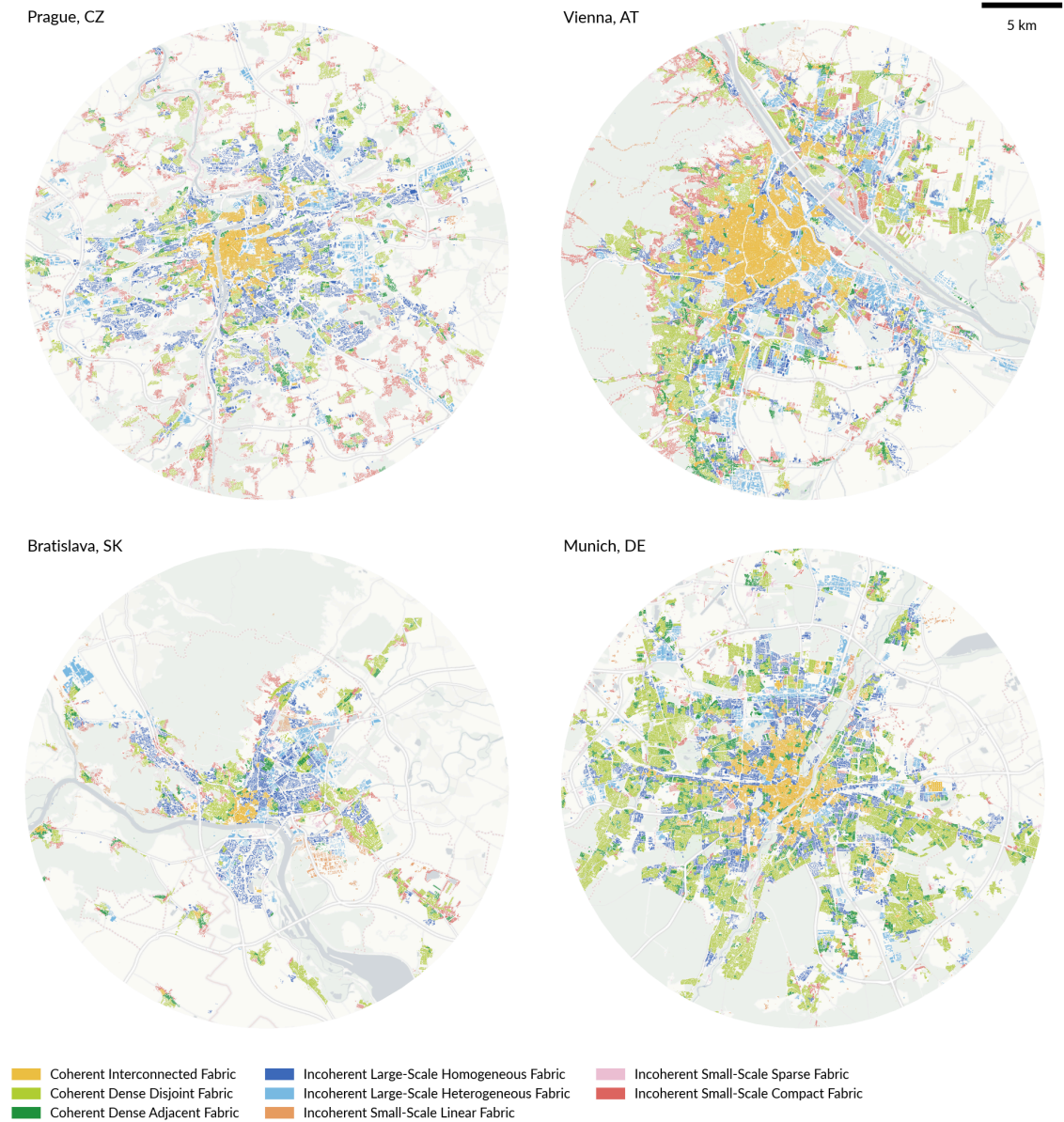

*Figure 5.* Taxonomy cut at level 3 illustrated in Prague, CZ, Vienna, AT, Bratislava, SK, and Munich, DE.

However, the taxonomy does not stop at the third level of bifurcation but can go arbitrarily deep. To illustrate that, Figure 6 shows deeper, unnamed taxonomy branches for the city of Prague, CZ. The first column shows that Prague at the third level of bifurcation, similarly to the cities and towns present in Figure 4. The next two columns represent deeper cuts. In the second column, the *Coherent Interconnected Fabric* is split into two clusters, one of which has a more organic street pattern and buildings which form larger blocks, while the other has a more regular grid pattern and the buildings form smaller blocks with fewer internal courtyards. The former cluster is further split in column three – smaller blocks with organic street patterns in one cluster, and larger buildings in interconnected blocks in the other. Similarly, the latter cluster with the grid pattern, is split based on street characteristics and courtyard presence.

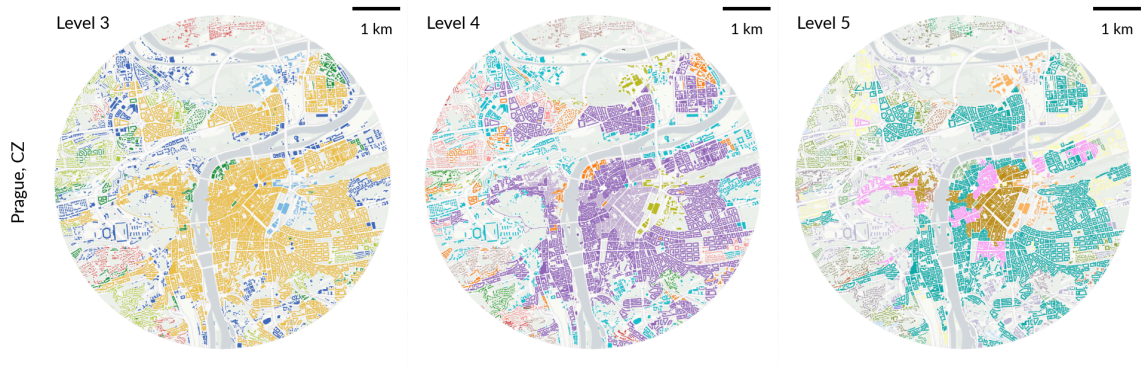

*Figure 6.* Deeper cuts through the taxonomy within Prague, CZ show that we can get arbitrarily granular classification all the way down to the level of individual morphotopes.

## 5.4 Spatial distribution of the results

Having a consistently derived, flexible taxonomy of this sort, covering multiple countries, enables multiple new types of analysis. While it is not our ambition to cover those within this publication, we analyse the spatial distribution of individual branches to showcase the scale, flexibility and potential applications of the taxonomy.

*Table 1.* Distribution of fabrics at level 3 per country in thousands of buildings. The value in brackets shows the proportion of buildings of a specific type to total buildings of a country.

|  | AT | CZ | DE | LT | PL | SK |
|---|---|---|---|---|---|---|
| *Incoherent Large-scale Homogeneous Fabric* | 132 (0.03) | 247 (0.06) | 1,911 (0.04) | 67 (0.03) | 780 (0.05) | 137 (0.04) |
| *Incoherent Large-scale Heterogeneous Fabric* | 97 (0.02) | 104 (0.03) | 1,811 (0.04) | 23 (0.01) | 289 (0.02) | 63 (0.02) |
| *Incoherent Small-scale Linear Fabric* | 223 (0.05) | 226 (0.05) | 1,937 (0.04) | 452 (0.22) | 4,156 (0.27) | 385 (0.11) |
| *Incoherent Small-scale Sparse Fabric* | 1,500 (0.34) | 1,102 (0.27) | 3,149 (0.06) | 621 (0.30) | 5,122 (0.33) | 234 (0.07) |
| *Incoherent Small-scale Compact Fabric* | 1,199 (0.27) | 1,078 (0.26) | 6,813 (0.13) | 506 (0.25) | 2,510 (0.16) | 730 (0.21) |
| *Coherent Interconnected Fabric* | 113 (0.03) | 81 (0.02) | 1,129 (0.02) | 2 (0.00) | 76 (0.01) | 25 (0.01) |
| *Coherent Dense Disjoint Fabric* | 788 (0.18) | 681 (0.17) | 28,918 (0.56) | 369 (0.18) | 2,024 (0.13) | 1,768 (0.51) |
| *Coherent Dense Adjacent Fabric* | 308 (0.07) | 609 (0.15) | 5,759 (0.11) | 8 (0.00) | 408 (0.03) | 91 (0.03) |

Table 1 shows the spatial distribution of the named clusters per country. Although every cluster is present in all countries, the distribution and density of specific urban forms is different. *Incoherent Large-Scale Homogeneous Fabric*, *Incoherent Large-scale Heterogeneous Fabric* , *Coherent Interconnected Fabric* are present in all countries in similar proportions. *Incoherent Small-scale Linear Fabric* is more prominent in Poland and Lithuania. *Incoherent Small-scale Sparse Fabric* is present in all countries at around 30%, apart from Germany and Slovakia. Inversely, around half the urban form in Germany and Slovakia is *Coherent Dense Disjoint Fabric*, while it is at around 15% in the other countries. *Incoherent Small-Scale Compact Fabric* is present everywhere at different proportions. *Coherent Dense Adjacent Fabric* is more prominent in Czechia, Austria and Germany.

Figure 7 shows these proportions on sub-national scale, using a rectangular grid of 50km edge size. What is shown further reiterates what is present in Table 1 but indicates that the abundance of individual types of built-up fabric varies also within countries, capturing their internal heterogeneity.

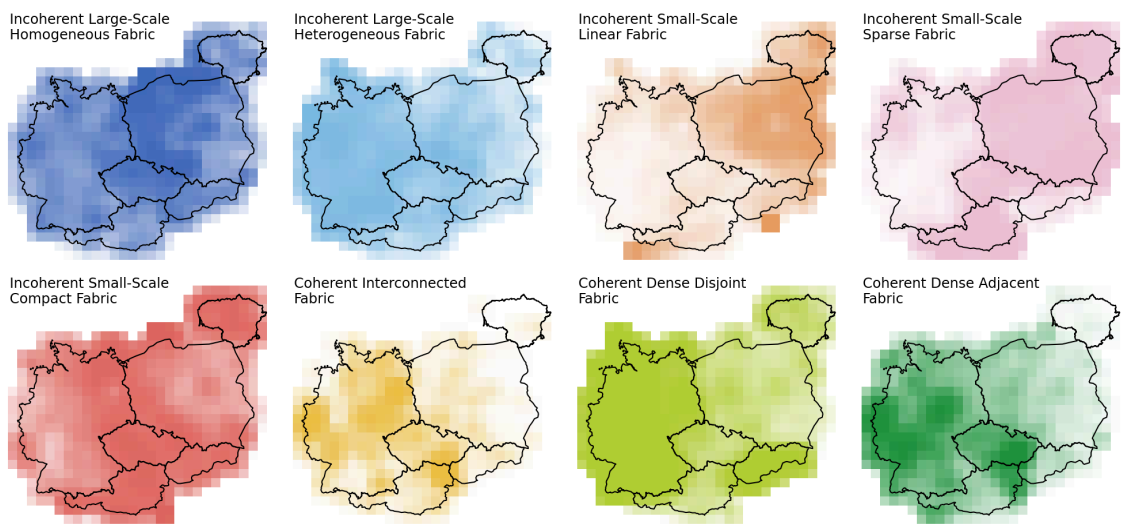

*Figure 7.* Relative abundance of branches at level 3 indicating presence of different fabrics in different geographical regions. The opacity of each color reflects the relative presence in each cell, compared to the observed maximum, showcasing spatial distribution of individual branches.

## 5.5 Comparisons

A quantitative comparison between HiMoC and existing data products can highlight the similarities and the uniqueness of the proposed classification method. We offer such a comparison in the Appendix G, based on co-occurrence, with three data products: CORINE Land Cover, Copernicus Urban Atlas and Local Climate Zones. In general, the branches of built-up fabric broadly follow the higher level patterns that are described by all three of data products. However, our branches tend to split their classes into different sub-classes, providing more information on internal structure of cities. One clear outcome of this comparison is that it seems that the large differences are due to different concepts being taken into account when making the classification. Given HiMoC is deeply embedded in morphological theory, it raises questions about the validity of of these classification, especially Local Climate Zones that are being used for this purpose, as a proxy for urban morphology outside of its original purpose (e.g. climate studies), as is being currently often used (Debray et al. 2025; Taubenböck et al. 2020).

## 6 Discussion & Conclusion

The *Hierarchical Morphotope Classification* represents another step in evolution of morphometric classification methods. One, that builds upon previous research (Fleischmann et al. 2022; Arribas-Bel and Fleischmann 2022; Araldi and Fusco 2024) and pushes its limits to allow capturing a much wider geographical extent (theoretically, we are limited only by data availability and consistency), while providing higher precision, significantly improved flexibility and a direct connection to morphological theory.

Previous research has been limited to individual cities and metropolitan areas (e.g. Fleischmann et al. 2022; Araldi and Fusco 2019), with occasional attempts to capture the fabric of an entire nation (Araldi and Fusco 2024; Fleischmann and Arribas-Bel 2022). HiMoC moves the limit of scalability beyond that, allowing consistent analysis of built-up fabric across multiple countries.

Its hierarchical structure allows a high degree of flexibility when using the classification within custom use cases. The resolution that is suitable for a top-level overview of large metropolitan areas might use 8 classes, while one that is required for detailed analysis of a single city might need to work with 80, or an advanced ecological analysis of presence of types of built-up fabric might rely on hundreds. All of which are derived in a consistent quantitative manner. Given that the top three levels come with a description and a pen-portrait, and every branch has a rich numeric profile, the flexible classification retains a high degree of interpretability at the same time.

The reliance on the morphotope concept and its operationalisation through the $SA^3$ regionalisation method comes with multiple benefits. First is an embedding of the core concept in Conzen's theory of urban morphology, ensuring that the method does not simply rely on some data output but that the output links to proved and verified ideas. Second is the fact that the regionalisation produces strictly contiguous boundaries, eliminating a degree of fuzziness present in methods with fewer spatial restrictions (Fleischmann et al. 2022; Arribas-Bel and Fleischmann 2022). Last is the adaptation to varying degrees of internal heterogeneity of built fabric. A method that treats heterogeneity uniformly typically gives too much space to locations where it is high, squashing the richness of relatively more homogeneous locations within the taxonomy.

While the taxonomic tree on top of morphotopes is conceptually similar to Conzen's hierarchy of morphological regions, we purposefully do not claim that it reflects it. The reason is a methodological difference between how morphological regions and their hierarchy are typically derived and how we derive our hierarchy. While this difference is still present when dealing with morphotopes, on such a small scale it is typically not visible. When going up the hierarchy, the difference between morphometric assessment of similarity and that of historical understanding of place grows. The two approaches start from the same idea but conceptualise it differently.

What the analysis of the taxonomy, as well as the morphometric profiling in Section E, shows is another key aspect of the work – the fact that the results were based on both building and street characteristics. If attributes derived from one or the other were concentrated at the top or at the bottom of the hierarchy, it would suggest that the method does not simply rely on some data output that only one set dominates the dissimilarity between the fabrics. However, that is not the case and both building and street patterns tend to affect the structure of the taxonomy at multiple levels. In fact, the top level separation is between, on the one hand, built fabric that closely follows the traditional relationship between buildings, plots and streets, and on the other, built fabric which breaks that relationship. Based on this result, classifications that focus only on streets or buildings in isolation are missing morphological information and potentially produce delineations at coarser scales of dissimilarity than is offered here.

Yet, the method comes with a set of limitations. At the moment, its applicability is restricted to areas where we have (relatively) consistent data inputs. Even though we rely only on 2-dimensional building footprint polygons and street networks, the building data is not always available and when it is, it might be inconsistent with INSPIRE (e.g. Microsoft Building Footprints) or even heterogeneous (e.g. OpenStreetMap and its derivatives like Overture Maps). Combining such data is not really possible as the morphometric attributes derived from different representations of buildings is not comparable. In the same vein, while the reliance on 2D data increases applicability, the method does not take into account height which might help distinguish certain branches, especially those containing modernist housing and utilitarian developments, where both showcase similar 2D patterns but are very different height-wise.

Furthermore, the stability of the taxonomic tree presented here is an open question. When extending the classification to cover additional countries, the tree will inevitably reshuffle. Having stable top branches would be desirable and strengthen the results of this work, but whether that is the case, we do not know at the moment.

To address this question, future work shall focus on expansion of the coverage of the classification. Many additional European countries offer data falling under the INSPIRE standard (e.g. France, Spain, The Netherlands, Belgium) and it is expected that these will be included in the classification in the upcoming years. Efforts like EUBUCCO (Milojevic-Dupont et al. 2023) might help with this progress by eliminating the need to fetch the data from individual countries but there still needs to be a robust framework for assessment of data consistency. Where the data is not available in the expected form, another branch of future work shall focus on gap filling methods using other types of data, be it satellite imagery or sub-optimal data coming from Overture Maps or other sources. At the same time, this classification reflects one point in time, without a temporal dimension. This is yet again mostly due to data availability, but we may look into a method based on satellite imagery that inherently comes as a time series to fill this dimension (Fleischmann and Arribas-Bel 2024).

Another fruitful research avenue shall focus on extraction of insights from the classification itself, relating it to the effect of political boundaries on the structure of settlements, environmental properties of different fabrics or relation of population groups and fabric types.

We need to better understand how our settlements are structured, what has formed them and how they affect the human population, environment or economy. None of this can be done without methods enabling analysis that are theory-driven, detailed, consistent and scalable. Given the complexity of urban morphology, we can expect that classification approaches, reducing the complexity to a small number of manageable and interpretable classes, will not fade away. HiMoC offers one such classification method, that aims to look at one specific slice of human settlements, their built form, and unpack it into a flexible hierarchical tree built on top of morphotopes, the smallest locality obtaining distinctive character among their neighbours (Conzen 1988). We believe that its complementarity with other existing large-scale classification methods paves the way for ubiquitous applicability beyond urban morphology and morphological analysis.

## 7 Code and data availability

All components of the work rely on open source software and open data, with the resulting code and data being openly available. Code, with the specification of a reproducible environment, is available at github.com/uscuni/urban_taxonomy and archived at doi.org/10.5281/zenodo.17105270. The classification is released as an open data product available at doi.org/10.5281/zenodo.17076283. The $SA^3$ functionality has been contributed to the open source package spopt (Feng et al. 2022), morphometric assessment to momepy (Fleischmann 2019), and features required to support those to libpysal (Rey et al. 2022), shapely (Gillies et al. 2025) and geopandas (Van den Bossche et al. 2025).

## 8 Acknowledgements

The authors kindly acknowledge funding by the Charles University's Primus programme through the project "Influence of Socioeconomic and Cultural Factors on Urban Structure in Central Europe", project reference PRIMUS/24/SCI/023. We would further like to thank to Claudia Baranzelli and Daniel Arribas-Bel for fruitful discussions helping shape the final outcome, to Rūta Marija Slavinskaitė, Milda Sutkaitytė, 4ct and Insitute for Planning and Devlopment Prague for their invaluable feedback on the classification and its applicability.

## 9 References

Abascal, Angela, Ignacio Rodríguez-Carreño, Sabine Vanhuysse, et al. 2022. "Identifying Degrees of Deprivation from Space Using Deep Learning and Morphological Spatial Analysis of Deprived Urban Areas." *Computers, Environment and Urban Systems* 95 (July): 101820. https://doi.org/10.1016/j.compenvurbsys.2022.101820.

Araldi, Alessandro, and Giovanni Fusco. 2019. "From the Street to the Metropolitan Region: Pedestrian Perspective in Urban Fabric Analysis." *Environment and Planning B: Urban Analytics and City Science* 46 (7): 1243–63. https://doi.org/10.1177/2399808319832612.

Araldi, Alessandro, and Giovanni Fusco. 2024. "Multi-Level Street-Based Analysis of the Urban Fabric: Developments for a Nationwide Taxonomy." *Geographical Analysis* 57 (2): 270–301. https://doi.org/10.1111/gean.12416.

Arat, Muzaffer Ali. 2023. "Urban Morphology and the Historico-Geographical Approach." In *Morphological Regionalization: Strengthening the Conzenian Method*, edited by Muzaffer Ali Arat. Springer Nature Switzerland. https://doi.org/10.1007/978-3-031-33509-9_2.

Arribas-Bel, Daniel, and Martin Fleischmann. 2022. "Spatial Signatures - Understanding (Urban) Spaces Through Form and Function." *Habitat International* 128 (October): 102641. https://doi.org/10.1016/j.habitat.2022.102641.

Basaraner, Melih, and Sinan Cetinkaya. 2017. "Performance of Shape Indices and Classification Schemes for Characterising Perceptual Shape Complexity of Building Footprints in GIS." *International Journal of Geographical Information Science* 31 (10): 1952–77. https://doi.org/10.1080/13658816.2017.1346257.

Batty, Michael. 2012. "Building a Science of Cities." *Cities* 29 (January): S9–16. https://doi.org/10.1016/j.cities.2011.11.008.

Biljecki, Filip, Yoong Shin Chow, and Kay Lee. 2023. "Quality of Crowdsourced Geospatial Building Information: A Global Assessment of OpenStreetMap Attributes." *Building and Environment* 237 (June): 110295. https://doi.org/10.1016/j.buildenv.2023.110295.

Bobkova, Evgeniya, Meta Berghauser Pont, and Lars Marcus. 2019. "Towards Analytical Typologies of Plot Systems: Quantitative Profile of Five European Cities." *Environment and Planning B: Urban Analytics and City Science*, October, 239980831988090. https://doi.org/10/ggbgsm.

Boeing, Geoff. 2017. "OSMnx: New Methods for Acquiring, Constructing, Analyzing, and Visualizing Complex Street Networks." *Computers, Environment and Urban Systems* 65 (September): 126–39. https://doi.org/10/gbvjxq.

Bongiorno, Christian, Yulun Zhou, Marta Kryven, et al. 2021. "Vector-Based Pedestrian Navigation in Cities." *Nature Computational Science* 1 (10): 678–85. https://doi.org/10.1038/s43588-021-00130-y.

Calafiore, Alessia, Krasen Samardzhiev, Francisco Rowe, Martin Fleischmann, and Daniel Arribas-Bel. 2023. "Inequalities in Experiencing Urban Functions. An Exploration of Human Digital (Geo-) Footprints." *Environment and Planning B: Urban Analytics and City Science*, 23998083231208507.

Caniggia, Gianfranco, and Gian Luigi Maffei. 1979. "Compositione Architettonica e Tipologia Edilizia. I Lettura Dell'edilizia Di Base." *Venice, Italy*, January.

Caniggia, Gianfranco, and Gian Luigi Maffei. 2001. *Architectural Composition and Building Typology: Interpreting Basic Building.* Vol. 176. Alinea Editrice.

Caruso, Geoffrey, Mohamed Hilal, and Isabelle Thomas. 2017. "Measuring Urban Forms from Inter-Building Distances: Combining MST Graphs with a Local Index of Spatial Association." *Landscape and Urban Planning* 163 (July): 80–89. https://doi.org/10.1016/j.landurbplan.2017.03.003.

Chen, Dongsheng, Yu Feng, Xun Li, Mingya Qu, Peng Luo, and Liqiu Meng. 2025. "Interpreting Core Forms of Urban Morphology Linked to Urban Functions with Explainable Graph Neural Network." *Computers, Environment and Urban Systems* 118 (June): 102267. https://doi.org/10.1016/j.compenvurbsys.2025.102267.

Chen, Wangyang, Abraham Noah Wu, and Filip Biljecki. 2021. "Classification of Urban Morphology with Deep Learning: Application on Urban Vitality." *Computers, Environment and Urban Systems* 90 (November): 101706. https://doi.org/10.1016/j.compenvurbsys.2021.101706.

Colaninno, Nicola, Josep Roca, and Karin Pfeffer. 2011. "An Automatic Classification of Urban Texture: Form and Compactness of Morphological Homogeneous Structures in Barcelona." *51st Congress of the European Regional Science Association*, January. https://www.econstor.eu/handle/10419/120085.

Conzen, Michael P. 2018. "Core Concepts in Town-Plan Analysis." In *Teaching Urban Morphology*, edited by Vítor Oliveira. Springer International Publishing. https://doi.org/10.1007/978-3-319-76126-8_8.

Conzen, Michael RG. 1988. "Morphogenesis, Morphological Regions and Secular Human Agency in the Historical Townscape as Exemplified by Ludlow." *Urban Historical Geography*.

Conzen, Michael Robert Gunter. 1960. "Alnwick, Northumberland: A Study in Town-Plan Analysis." *Transactions and Papers (Institute of British Geographers)*, no. 27: iii–122.

Conzen, Michael Robert Gunter. 1975. "Geography and Townscape Conservation." *Anglo-German Symposium in Applied Geography*, 95–102.

Craglia, Max, and Alessandro Annoni. 2007. "INSPIRE: An Innovative Approach to the Development of Spatial Data Infrastructures in Europe." *Research and Theory in Advancing Spatial Data Infrastructure Concepts*, 93–105.

De Sabbata, Stefano, Andrea Ballatore, Pengyuan Liu, and Nicholas Tate. 2023. *Learning Urban Form Through Unsupervised Graph-Convolutional Neural Networks.* August. https://doi.org/10.17605/OSF.IO/H2AWQ'].

Debray, Henri, Matthias Gassilloud, Richard Lemoine-Rodríguez, Michael Wurm, Xiaoxiang Zhu, and Hannes Taubenböck. 2025. "Universal Patterns of Intra-Urban Morphology: Defining a Global Typology of the Urban Fabric Using Unsupervised Clustering." *International Journal of Applied Earth Observation and Geoinformation* 141 (July): 104610. https://doi.org/10.1016/j.jag.2025.104610.

Demuzere, M., J. Kittner, A. Martilli, et al. 2022. "A Global Map of Local Climate Zones to Support Earth System Modelling and Urban-Scale Environmental Science." *Earth System Science Data* 14 (8): 3835–73. https://doi.org/10.5194/essd-14-3835-2022.

Dibble, Jacob, Alexios Prelorendjos, Ombretta Romice, et al. 2015. "Urban Morphometrics: Towards a Science of Urban Evolution." *arXiv.org* physics.soc-ph (June). http://arxiv.org/abs/1506.04875v2.

Dibble, Jacob, Alexios Prelorendjos, Ombretta Romice, et al. 2019. "On the Origin of Spaces: Morphometric Foundations of Urban Form Evolution." *Environment and Planning B: Urban Analytics and City Science* 46 (4): 707–30.

Domingo, Darío, Jasper van Vliet, and Anna M. Hersperger. 2023. "Long-Term Changes in 3D Urban Form in Four Spanish Cities." *Landscape and Urban Planning* 230: 104624. https://doi.org/https://doi.org/10.1016/j.landurbplan.2022.104624.

European Environment Agency. 1990. *CORINE Land Cover.* January, 1–163. http://www.eea.europa.eu/publications/COR0-landcover/page001.html.

European Environment Agency. 2020. *Urban Atlas Land Cover/Land Use 2018 (Vector), Europe, 6-Yearly, Jul. 2021.* European Environment Agency. https://doi.org/10.2909/FB4DFFA1-6CEB-4CC0-8372-1ED354C285E6.

Fedchenko, Irina. 2023. "Morphological Studies as a Prerequisite for the Development of City Regulations." January, 293–99. https://doi.org/10.55060/s.atssh.221230.039.

Feliciotti, Alessandra. 2018. "RESILIENCE AND URBAN DESIGN: A SYSTEMS APPROACH TO THE STUDY OF RESILIENCE IN URBAN FORM." PhD thesis, University of Strathclyde.

Feng, Xin, Germano Barcelos, James D. Gaboardi, et al. 2022. "Spopt: A Python Package for Solving Spatial Optimization Problems in PySAL." *Journal of Open Source Software* 7 (74): 3330. https://doi.org/10.21105/joss.03330.

Fleischmann, Martin. 2019. "Momepy: Urban Morphology Measuring Toolkit." *Journal of Open Source Software* 4 (43): 1807. https://doi.org/10.21105/joss.01807.

Fleischmann, Martin, and Daniel Arribas-Bel. 2022. "Geographical Characterisation of British Urban Form and Function Using the Spatial Signatures Framework." *Scientific Data* 9 (1): 546.

Fleischmann, Martin, and Daniel Arribas-Bel. 2024. "Decoding (Urban) Form and Function Using Spatially Explicit Deep Learning." *Computers, Environment and Urban Systems* 112 (September): 102147. https://doi.org/10.1016/j.compenvurbsys.2024.102147.

Fleischmann, Martin, Alessandra Feliciotti, Ombretta Romice, and Sergio Porta. 2020. "Morphological Tessellation as a Way of Partitioning Space: Improving Consistency in Urban Morphology at the Plot Scale." *Computers, Environment and Urban Systems* 80 (January): 101441. https://doi.org/10.1016/j.compenvurbsys.2019.101441.

Fleischmann, Martin, Alessandra Feliciotti, Ombretta Romice, and Sergio Porta. 2022. "Methodological Foundation of a Numerical Taxonomy of Urban Form." *Environment and Planning B: Urban Analytics and City Science* 49 (4): 1283–99. https://doi.org/10/gnth7q.

Fleischmann, Martin, and Anastassia Vybornova. 2024. "A Shape-Based Heuristic for the Detection of Urban Block Artifacts in Street Networks." *Journal of Spatial Information Science* 28 (June): 75–102. https://doi.org/10.5311/JOSIS.2024.28.319.

Fleischmann, Martin, Anastassia Vybornova, James D. Gaboardi, Anna Brázdová, and Daniela Dančejová. 2025. *Adaptive Continuity-Preserving Simplification of Street Networks.* arXiv:2504.16198. arXiv. https://doi.org/10.48550/arXiv.2504.16198.

Gil, Jorge, Nuno Montenegro, J N Beirão, and J P Duarte. 2012. "On the Discovery of Urban Typologies: Data Mining the Multi-Dimensional Character of Neighbourhoods." *Urban Morphology* 16 (1): 27–40. http://papers.cumincad.org/cgi-bin/works/Show?ecaade2009_148.

Gillies, Sean, Casper van der Wel, Joris Van den Bossche, et al. 2025. *Shapely.* Version 2.1.1. Zenodo, released May. https://doi.org/10.5281/zenodo.15463269.

Guyot, Madeleine, Alessandro Araldi, Giovanni Fusco, and Isabelle Thomas. 2021. "The Urban Form of Brussels from the Street Perspective: The Role of Vegetation in the Definition of the Urban Fabric." *Landscape and Urban Planning* 205 (January): 103947. https://doi.org/10/ghf96c.

Hamaina, Rachid, Thomas Leduc, and Guillaume Moreau. 2012. "Towards Urban Fabrics Characterization Based on Buildings Footprints." In *Bridging the Geographic Information Sciences*, vol. 2, 2. Springer, Berlin, Heidelberg. https://doi.org/10.1007/978-3-642-29063-3_18.

Hasanzadeh, Kamyar, Marketta Kyttä, and Greg Brown. 2019. "Beyond Housing Preferences: Urban Structure and Actualisation of Residential Area Preferences." *Urban Science* 3 (1): 21. https://doi.org/10.3390/urbansci3010021.

Hijazi, Ihab, Xin Li, Reinhard Koenig, et al. 2016. "Measuring the Homogeneity of Urban Fabric Using 2D Geometry Data." *Environment and Planning B: Planning and Design*, January, 1–25. https://doi.org/10.1177/0265813516659070.

Huang, Jingnan, X X Lu, and Jefferey M Sellers. 2007. "A Global Comparative Analysis of Urban Form: Applying Spatial Metrics and Remote Sensing." *Landscape and Urban Planning* 82 (4): 184–97. https://doi.org/10.1016/j.landurbplan.2007.02.010.

Jochem, Warren C, Douglas R Leasure, Oliver Pannell, Heather R Chamberlain, Patricia Jones, and Andrew J Tatem. 2020. "Classifying Settlement Types from Multi-Scale Spatial Patterns of Building Footprints." *Environment and Planning B: Urban Analytics and City Science*, May, 239980832092120. https://doi.org/10/ggtsbn.

Levy, Albert. 1999. "Urban Morphology and the Problem of the Modern Urban Fabric: Some Questions for Research." *Urban Morphology* 3 (January): 79–85.

Lind, Pedro G., Marta C. González, and Hans J. Herrmann. 2005. "Cycles and Clustering in Bipartite Networks." *Physical Review E* 72 (5): 056127. https://doi.org/10/c6m9xd.

Liu, Ziyu, and Yacheng Song. 2025. "Unsupervised Plot Morphology Classification via Graph Attention Networks: Evidence from Nanjing's Walled City." *Land* 14 (7): 1469. https://doi.org/10.3390/land14071469.

Lynch, Kevin. 1960. *The Image of the City.* Vol. 11. MIT press.

Marwal, Aviral, and Elisabete A. Silva. 2023. "Exploring Residential Built-up Form Typologies in Delhi: A Grid-Based Clustering Approach Towards Sustainable Urbanisation." *Npj Urban Sustainability* 3 (1): 40. https://doi.org/10.1038/s42949-023-00112-1.

McInnes, Leland, John Healy, and Steve Astels. 2017. "Hdbscan: Hierarchical Density Based Clustering." *Journal of Open Source Software* 2 (11): 205. https://doi.org/10/ggfp85.

Milojevic-Dupont, Nikola, Nicolai Hans, Lynn H Kaack, et al. 2020. "Learning from Urban Form to Predict Building Heights." *Plos One* 15 (12): e0242010. https://doi.org/10/ghn8rc.

Milojevic-Dupont, Nikola, Felix Wagner, Florian Nachtigall, et al. 2023. "EUBUCCO V0.1: European Building Stock Characteristics in a Common and Open Database for 200+ Million Individual Buildings." *Scientific Data* 10 (1): 147. https://doi.org/10.1038/s41597-023-02040-2.

Moudon, Anne Vernez. 1997. "Urban Morphology as an Emerging Interdisciplinary Field." *Urban Morphology* 1 (1): 3–10.

Muratori, Saverio. 1959. "Studi Per Una Operante Storia Urbana Di Venezia." *Palladio. Rivista Di Storia Dell'architettura* 1959: 1–113.

Oliveira, Vítor. 2019. "An Historico-Geographical Theory of Urban Form." *Journal of Urbanism: International Research on Placemaking and Urban Sustainability* 12 (4): 412–32. https://doi.org/10/gg639n.

Oliveira, Vitor, and Sergio Porta. 2025. "Quantitative and Qualitative Analysis in Urban Morphology: Systematic Legacy and Latest Developments." *Proceedings of the Institution of Civil Engineers - Urban Design and Planning* 178 (2): 75–87. https://doi.org/10.1680/jurdp.24.00047.

Porta, Sergio, Luciano Da Fontoura Costa, Eugenio Morello, Emanuele Strano, Alessandro Venerandi, and Ombretta Romice. 2011. *Plot-Based Urbanism and Urban Morphometrics: Measuring the Evolution of Blocks, Street Fronts and Plots in Cities.* January.

Rey, Sergio J., Luc Anselin, Pedro Amaral, et al. 2022. "The PySAL Ecosystem: Philosophy and Implementation." *Geographical Analysis* 54 (3): 467–87. https://doi.org/10.1111/gean.12276.

Schirmer, Patrick Michael, and Kay W Axhausen. 2015. "A Multiscale Classification of Urban Morphology." *Journal of Transport and Land Use* 9 (1): 101–30. https://doi.org/10.5198/jtlu.2015.667.

Singleton, Alex, and Daniel Arribas-Bel. 2019. "Geographic Data Science." *Geographical Analysis* 53 (1): 61–75. https://doi.org/10.1111/gean.12194.

Slater, T. R. 1981. "The Analysis of Burgage Patterns in Medieval Towns." *Area* 13 (3): 211–16. http://www.jstor.org/stable/20001722.

Smith, D, and A Crooks. 2010. *From Buildings to Cities: Techniques for the Multi-Scale Analysis of Urban Form and Function.* January. http://discovery.ucl.ac.uk/20360/.

Song, Yan, and Gerrit-Jan Knaap. 2007. "Quantitative Classification of Neighbourhoods: The Neighbourhoods of New Single-Family Homes in the Portland Metropolitan Area." *Journal of Urban Design* 12 (1): 1–24. https://doi.org/10.1080/13574800601072640.

Steiniger, Stefan, Tilman Lange, Dirk Burghardt, and Robert Weibel. 2008. "An Approach for the Classification of Urban Building Structures Based on Discriminant Analysis Techniques." *Transactions in GIS* 12 (1): 31–59. https://doi.org/10.1111/j.1467-9671.2008.01085.x.

Stewart, I D, and T R Oke. 2012. "Local Climate Zones for Urban Temperature Studies." *Bulletin of the American Meteorological Society* 93 (12): 1879–900. https://doi.org/10.1175/BAMS-D-11-00019.1.

Taubenböck, H., H. Debray, C. Qiu, M. Schmitt, Y. Wang, and X. X. Zhu. 2020. "Seven City Types Representing Morphologic Configurations of Cities Across the Globe." *Cities* 105 (October): 102814. https://doi.org/10/gg2jv4.

Thomas, Isabelle, Pierre Frankhauser, Benoit Frenay, Michel Verleysen, and S M Samos-Matisse. 2010. "Clustering Patterns of Urban Built-up Areas with Curves of Fractal Scaling Behaviour." *Environment and Planning B: Planning and Design* 37 (5): 942–54. https://doi.org/10.1068/b36039.

Van den Bossche, Joris, Kelsey Jordahl, Martin Fleischmann, et al. 2025. *Geopandas/Geopandas: V1.1.1.* Version v1.1.1. Zenodo, released June. https://doi.org/10.5281/zenodo.15750510.

Vanderhaegen, Sven, and Frank Canters. 2017. "Mapping Urban Form and Function at City Block Level Using Spatial Metrics." *Landscape and Urban Planning* 167 (November): 399–409. https://doi.org/10.1016/j.landurbplan.2017.05.023.

Venerandi, Alessandro, Alessandra Feliciotti, Safoora Mokhtarzadeh, Maryam Taefnia, Ombretta Romice, and Sergio Porta. 2024. "Urban Form and Socioeconomic Deprivation in Isfahan: An Urban MorphoMetric Approach." *Environment and Planning B: Urban Analytics and City Science* 51 (9): 2232–48. https://doi.org/10.1177/23998083241245491.

Wang, Jiong, Martin Fleischmann, Alessandro Venerandi, Ombretta Romice, Monika Kuffer, and Sergio Porta. 2023. "EO+ Morphometrics: Understanding Cities Through Urban Morphology at Large Scale." *Landscape and Urban Planning* 233: 104691. https://doi.org/10.1016/j.landurbplan.2023.104691.

Wang, Kechao, Tingting He, Wu Xiao, and Runjia Yang. 2024. "Projections of Future Spatiotemporal Urban 3D Expansion in China Under Shared Socioeconomic Pathways." *Landscape and Urban Planning* 247: 105043. https://doi.org/https://doi.org/10.1016/j.landurbplan.2024.105043.

Whitehand, Jeremy WR. 1967. "Fringe Belts: A Neglected Aspect of Urban Geography." *Transactions of the Institute of British Geographers*, 223–33.

Wu, Zhifeng, Ying Wang, and Yin Ren. 2025. "Optimizing Green Space-Building Landscape Characteristics of Key Urban Functional Zones for Comprehensive Thermal Environment Mitigation." *Landscape and Urban Planning* 257: 105314. https://doi.org/https://doi.org/10.1016/j.landurbplan.2025.105314.

Wurm, Michael, Hannes Taubenbock, and Stefan Dech. 2010. "Quantification of Urban Structure on Building Block Level Utilizing Multisensoral Remote Sensing Data." 7831 (October): 78310H. https://doi.org/10.1117/12.864930.

Yu, Tengfei, Birgit S Sützl, and Maarten van Reeuwijk. 2023. "Urban Neighbourhood Classification and Multi-Scale Heterogeneity Analysis of Greater London." *Environment and Planning B: Urban Analytics and City Science* 50 (6): 1534–58. https://doi.org/10.1177/23998083221140890.

Zhu, Sijie, Yanxia Li, Shen Wei, et al. 2022. "The Impact of Urban Vegetation Morphology on Urban Building Energy Consumption During Summer and Winter Seasons in Nanjing, China." *Landscape and Urban Planning* 228: 104576. https://doi.org/https://doi.org/10.1016/j.landurbplan.2022.104576.

# A. Morphometric Characters

Characteristics describing the interactions of these elements, and the elements themselves are calculated at three scales: small - covering only aspects of the element; medium - covering aspects of the element and neighbouring elements and large - covering neighbouring elements up to five topological neighbours.

1. **Area of a building** is denoted as

(1) $a_{blg}$

and defined as an area covered by a building footprint in m² .

2. **Perimeter of a building** is denoted as

(2) $p_{blg}$

and defined as the sum of lengths of the building exterior walls in m.

3. **Courtyard area of a building** is denoted as

(3) $a_{blg_c}$

and defined as the sum of areas of interior holes in footprint polygons in m².

4. **Circular compactness of a building** is denoted as

(4) $CCo_{blg} = \frac{a_{blg}}{a_{blgC}}$

where $a_{blgC}$ is an area of minimal enclosing circle. It captures the relation of building footprint shape to its minimal enclosing circle, illustrating the similarity of shape and circle (Dibble et al. 2015).

5. **Corners of a building** is denoted as

(5) $Cor_{blg} = \sum_{i=1}^{n} c_{blg}$

where $c_{blg}$ is defined as a vertex of building exterior shape with an angle between adjacent line segments $\leq 170$ degrees. It uses only external shape (`shapely.geometry.exterior`), courtyards are not included. Character is adapted from (Steiniger et al. 2008) to exclude non-corner-like vertices.

6. **Squareness of a building** is denoted as

(6) $Squ_{blg} = \frac{\sum_{i=1}^{n} D_{c_{blg_i}}}{n}$

where $D$ is the deviation of angle of corner $c_{blg_i}$ from 90 degrees and $n$ is a number of corners.

7. **Equivalent rectangular index of a building** is denoted as

(7) $ERI_{blg} = \sqrt{\frac{a_{blg}}{a_{blgB}} \times \frac{p_{blgB}}{p_{blg}}}$

where $a_{blgB}$ is an area of a minimal rotated bounding rectangle of a building (MBR) footprint and $p_{blgB}$ its perimeter of MBR. It is a measure of shape complexity identified by Basaraner and Cetinkaya (2017) as the shape characters with the best performance.

8. **Elongation of a building** is denoted as

(8) $Elo_{blg} = \frac{l_{blgB}}{w_{blgB}}$

where $l_{blgB}$ is length of MBR and $w_{blgB}$ is width of MBR. It captures the ratio of shorter to the longer dimension of MBR to indirectly capture the deviation of the shape from a square (Schirmer and Axhausen 2015).

9. **Longest axis length of a tessellation cell** is denoted as

(9) $LAL_{cell} = d_{cellC}$

where $d_{cellC}$ is a diameter of the minimal circumscribed circle around the tessellation cell polygon. The axis itself does not have to be fully within the polygon. It could be seen as a proxy of plot depth for tessellation-based analysis.

10. **Area of a tessellation cell** is denoted as

(10) $a_{cell}$

and defined as an area covered by a tessellation cell footprint in m².

11. **Circular compactness of a tessellation cell** is denoted as

(11) $CCo_{cell} = \frac{a_{cell}}{a_{cellC}}$

where $a_{cellC}$ is an area of minimal enclosing circle. It captures the relation of tessellation cell footprint shape to its minimal enclosing circle, illustrating the similarity of shape and circle.

12. **Equivalent rectangular index of a tessellation cell** is denoted as

(12) $ERI_{cell} = \sqrt{\frac{a_{cell}}{a_{cellB}}} \times \frac{p_{cellB}}{p_{cell}}$

where $a_{cellB}$ is an area of the minimal rotated bounding rectangle of a tessellation cell (MBR) footprint and $p_{cellB}$ its perimeter of MBR. It is a measure of shape complexity identified by Basaraner and Cetinkaya (2017) as a shape character of the best performance.

13. **Coverage area ratio of a tessellation cell** is denoted as

(13) $CAR_{cell} = \frac{a_{blg}}{a_{cell}}$

where $a_{blg}$ is an area of a building and $a_{cell}$ is an area of related tessellation cell (Schirmer and Axhausen 2015). Coverage area ratio (CAR) is one of the commonly used characters capturing *intensity* of development. However, the definitions vary based on the spatial unit.

14. **Length of a street segment** is denoted as

(14) $l_{edg}$

and defined as a length of a `LineString` geometry in metres (Dibble et al. 2015; Gil et al. 2012).

15. **Width of a street profile** is denoted as

(15) $w_{sp} = \frac{1}{n}\left(\sum_{i=1}^{n} w_i\right)$

where $w_i$ is width of a street section i. The algorithm generates street sections every 3 meters alongside the street segment, and measures mean value. In the case of the open-ended street, 50 metres is used as a perception-based proximity limit (Araldi and Fusco 2019).

16. **Openness of a street profile** is denoted as

(16) $Ope_{sp} = 1 - \frac{\sum hit}{2 \sum sec}$

where $\sum hit$ is a sum of section lines (left and right sides separately) intersecting buildings and $\sum sec$ total number of street sections. The algorithm generates street sections every 3 meters alongside the street segment.

17. **Width deviation of a street profile** is denoted as

(17) $wDev_{sp} = \sqrt{\frac{1}{n}\sum_{i=1}^{n}\left(w_i - w_{sp}\right)^2}$

where $w_i$ is width of a street section i and $w_{sp}$ is mean width. The algorithm generates street sections every 3 meters alongside the street segment.

18. **Linearity of a street segment** is denoted as

(18) $Lin_{edg} = \frac{l_{eucl}}{l_{edg}}$

where $l_{eucl}$ is Euclidean distance between endpoints of a street segment and $l_{edg}$ is a street segment length. It captures the deviation of a segment shape from a straight line. It is adapted from Araldi and Fusco (2019).

19. **Area covered by a street segment** is denoted as

(19) $a_{edg} = \sum_{i=1}^{n} a_{cell_i}$

where $a_{cell_i}$ is an area of tessellation cell $i$ belonging to the street segment. It captures the area which is likely served by each segment.

20. **Buildings per meter of a street segment** is denoted as

(20) $BpM_{edg} = \frac{\sum blg}{l_{edg}}$

where $\sum blg$ is a number of buildings belonging to a street segment and $l_{edg}$ is a length of a street segment. It reflects the granularity of development along each segment.

21. **Area covered by a street node** is denoted as

(21) $a_{node} = \sum_{i=1}^{n} a_{cell_i}$

where $a_{cell_i}$ is an area of tessellation cell $i$ belonging to the street node. It captures the area which is likely served by each node.

22. **Shared walls ratio of adjacent buildings** is denoted as

(22) $SWR_{blg} = \frac{p_{blg_{shared}}}{p_{blg}}$

where $p_{blg_{shared}}$ is a length of a perimeter shared with adjacent buildings and $p_{blg}$ is a perimeter of a building. It captures the amount of wall space facing the open space (Hamaina et al. 2012).

23. **Mean distance to neighbouring buildings** is denoted as

(23) $NDi_{blg} = \frac{1}{n} \sum_{i=1}^{n} d_{blg,blg_i}$

where $d_{blg,blg_i}$ is a distance between building and building $i$ on a neighbouring tessellation cell. It is adapted from Hijazi et al. (2016). It captures the average proximity to other buildings.

24. **Weighted neighbours of a tessellation cell** is denoted as

(24) $WNe_{cell} = \frac{\sum cell_n}{p_{cell}}$

where $\sum cell_n$ is a number of cell neighbours and $p_{cell}$ is a perimeter of a cell. It reflects granularity of morphological tessellation.

25. **Area covered by neighbouring cells** is denoted as

(25) $a_{cell_n} = \sum_{i=1}^{n} a_{cell_i}$

where $a_{cell_i}$ is area of tessellation cell $i$ within topological distance 1. It captures the scale of morphological tessellation.

26. **Reached area by neighbouring segments** is denoted as

(26) $a_{edg_n} = \sum_{i=1}^{n} a_{edg_i}$

where $a_{edg_i}$ is an area covered by a street segment $i$ within topological distance 1. It captures an accessible area.

27. **Degree of a street node** is denoted as

(27) $deg_{node_i} = \sum_j edg_{ij}$

where $edg_{ij}$ is an edge of a street network between node $i$ and node $j$. It reflects the basic degree centrality.

28. **Mean distance to neighbouring nodes from a street node** is denoted as

(28) $MDi_{node} = \frac{1}{n} \sum_{i=1}^{n} d_{node,node_i}$

where $d_{node,node_i}$ is a distance between node and node $i$ within topological distance 1. It captures the average proximity to other nodes.

29. **Reached cells by neighbouring nodes** is denoted as

(29) $RC_{node_n} = \sum_{i=1}^{n} cells_{node_i}$

where $cells_{node_i}$ is number of tessellation cells on node $i$ within topological distance 1. It captures accessible granularity.

30. **Reached area by neighbouring nodes** is denoted as

(30) $a_{node_n} = \sum_{i=1}^{n} a_{node_i}$

where $a_{node_i}$ is an area covered by a street node $i$ within topological distance 1. It captures an accessible area.

31. **Number of courtyards of adjacent buildings** is denoted as

(31) $NCo_{blg_{adj}}$

where $NCo_{blg_{adj}}$ is a number of interior rings of a polygon composed of footprints of adjacent buildings (Schirmer and Axhausen 2015).

32. **Perimeter wall length of adjacent buildings** is denoted as

(32) $p_{blg_{adj}}$

where $p_{blg_{adj}}$ is a length of an exterior ring of a polygon composed of footprints of adjacent buildings.

33. **Mean inter-building distance between neighbouring buildings** is denoted as

(33) $IBD_{blg} = \frac{1}{n} \sum_{i=1}^{n} d_{blg,blg_i}$

where $d_{blg,blg_i}$ is a distance between building and building $i$ on a tessellation cell within topological distance 3. It is adapted from Caruso et al. (2017). It captures the average proximity between buildings.

34. **Building adjacency of neighbouring buildings** is denoted as

(34) $BuA_{blg} = \frac{\sum blg_{adj}}{\sum blg}$

where $\sum blg_{adj}$ is a number of joined built-up structures within topological distance three and $\sum blg$ is a number of buildings within topological distance 3. It is adapted from Vanderhaegen and Canters (2017).

35. **Weighted reached blocks of neighbouring tessellation cells** is denoted as

(35) $WRB_{cell} = \frac{\sum blk}{\sum_{i=1}^{n} a_{cell_i}}$

where $\sum blk$ is a number of blocks within topological distance three and $a_{cell_i}$ is an area of tessellation cell $i$ within topological distance three.

36. **Local meshedness of a street network** is denoted as

(36) $Mes_{node} = \frac{e-v+1}{2v-5}$

where $e$ is a number of edges in a subgraph, and $v$ is the number of nodes in a subgraph (Feliciotti 2018). A subgraph is defined as a network within topological distance five around a node.

37. **Mean segment length of a street network** is denoted as

(37) $MSL_{edg} = \frac{1}{n} \sum_{i=1}^{n} l_{edg_i}$

where $l_{edg_i}$ is a length of a street segment $i$ within a topological distance 3 around a segment.

38. **Cul-de-sac length of a street network** is denoted as

(38) $CDL_{node} = \sum_{i=1}^{n} l_{edg_i}$, if $edg_i$ is cul-de-sac

where $l_{edg_i}$ is a length of a street segment $i$ within a topological distance 3 around a node.

39. **Reached cells by street network segments** is denoted as

(39) $RC_{edg} = \sum_{i=1}^{n} cells_{edg_i}$

where $cells_{edg_i}$ is number of tessellation cells on segment $i$ within topological distance 3. It captures accessible granularity.

40. **Node density of a street network** is denoted as

(40) $D_{node} = \frac{\sum node}{\sum_{i=1}^{n} l_{edg_i}}$

where $\sum node$ is a number of nodes within a subgraph and $l_{edg_i}$ is a length of a segment $i$ within a subgraph. A subgraph is defined as a network within topological distance five around a node.

41. **Reached cells by street network nodes** is denoted as

(41) $RC_{node_{net}} = \sum_{i=1}^{n} cells_{node_i}$

where $cells_{node_i}$ is number of tessellation cells on node $i$ within topological distance 3. It captures accessible granularity.

42. **Reached area by street network nodes** is denoted as

(42) $a_{node_{net}} = \sum_{i=1}^{n} a_{node_i}$

where $a_{node_i}$ is an area covered by a street node $i$ within topological distance 3. It captures an accessible area.

43. **Proportion of cul-de-sacs within a street network** is denoted as

(43) $pCD_{node} = \frac{\sum_{i=1}^{n} node_i, \ if \ deg_{node_i}=1}{\sum_{i=1}^{n} node_i}$

where $node_i$ is a node whiting topological distance five around a node. Adapted from (Boeing 2017).

44. **Proportion of 3-way intersections within a street network** is denoted as

(44) $p3W_{node} = \frac{\sum_{i=1}^{n} node_i, \ if \ deg_{node_i}=3}{\sum_{i=1}^{n} node_i}$

where $node_i$ is a node whiting topological distance five around a node. Adapted from (Boeing 2017).

45. **Proportion of 4-way intersections within a street network** is denoted as

(45) $p4W_{node} = \frac{\sum_{i=1}^{n} node_i, \ if \ deg_{node_i}=4}{\sum_{i=1}^{n} node_i}$

where $node_i$ is a node whiting topological distance five around a node. Adapted from (Boeing 2017).

46. **Weighted node density of a street network** is denoted as

(46) $wD_{node} = \frac{\sum_{i=1}^{n} deg_{node_i}-1}{\sum_{i=1}^{n} l_{edg_i}}$

where $deg_{node_i}$ is a degree of a node $i$ within a subgraph and $l_{edg_i}$ is a length of a segment $i$ within a subgraph. A subgraph is defined as a network within topological distance five around a node.

47. **Local closeness centrality of a street network** is denoted as

(47) $lCC_{node} = \frac{n-1}{\sum_{v=1}^{n-1} d(v,u)}$

where $d(v,u)$ is the shortest-path distance between $v$ and $u$, and $n$ is the number of nodes within a subgraph. A subgraph is defined as a network within topological distance five around a node.

48. **Square clustering of a street network** is denoted as

(48) $sCl_{node} = \frac{\sum_{u=1}^{k_v} \sum_{w=u+1}^{k_v} q_v(u,w)}{\sum_{u=1}^{k_v} \sum_{w=u+1}^{k_v} [a_v(u,w) + q_v(u,w)]}$

where $q_v(u,w)$ are the number of common neighbours of $u$ and $w$ other than $v$ (ie squares), and $a_v(u,w) = (k_u - (1 + q_v(u,w) + \theta_{uv}))(k_w - (1 + q_v(u,w) + \theta_{uw}))$, where $\theta_{uw} = 1$ if $u$ and $w$ are connected and 0 otherwise (Lind et al. 2005).

49. **Connected buildings count** is denoted as

(49) $c_{blg}$

and defined as number of buildings directly adjacent to the target building.

50. **Connected buildings area** is denoted as

(50) $a_{cblg}$

and defined as total area of all buildings directly adjacent to the target building.

51. **Connected buildings perimeter** is denoted as

(51) $p_{cblg}$

and defined as total perimeter of all buildings directly adjacent to the target building.

52. **Connected buildings elongation** is denoted as

(52) $mibElo_{cblg} = Elo(cblg)$

where $cblg$ are all buildings adjacent to the target building and $Elo$ is the elongation formula defined previously.

53. **Connected buildings elongation** is denoted as

(53) $mibERI_{cblg} = ERI(cblg)$

where $cblg$ are all buildings adjacent to the target building and $ERI$ is the elongation formula defined previously.

54. **Connected buildings circular compactness** is denoted as

(54) $mibCCo_{cblg} = CCo(cblg)$

where $cblg$ are all buildings adjacent to the target building and $CCo$ is the elongation formula defined previously.

55. **Connected buildings longest axis length** is denoted as

(55) $mibLAL_{cblg} = LAL(cblg)$

where $cblg$ are all buildings adjacent to the target building and $LAL$ is the elongation formula defined previously.

56. **Connected buildings facade ratio** is denoted as

(56) $mibFR_{cblg} = \frac{mibAre_{cblg}}{mibPer_{cblg}}$

where $cblg$ are all buildings adjacent to the target building and $mibAre$ and $mibPer$ are the formulas defined previously.

57. **Connected buildings square compactness** is denoted as

(57) $mibSCo_{cblg} = \left(\frac{4\sqrt{mibAre_{cblg}}}{mibPer_{cblg}}\right)^2$

where $cblg$ are all buildings adjacent to the target building and $mibAre$ and $mibPer$ are the formulas defined previously.

58. **Deviation of building area in tessellation neighbourhood** is denoted as

(58) $micBAD_{cell}$

and is defined as the standard deviation in the areas of all buildings within tessellation cells, directly adjacent to the target tesellation cell.

59. **Deviation of building area in node-attached buildings** is denoted as

(59) $midBAD_{node}$

and is defined as the standard deviation in the areas of all buildings attached to the target node.

There are three additional indicator variable calculated per morphotope - "Likely Occupied Area", "Area of the largest ten connected structures" and "Perimeter of the largest ten connected structures".

The "Likely Occupied Area" variable aims to identify morphotopes that mainly contain industrial buildings which tend to be blocky and not built to optimise light exposure. This variable is required due to inconsistencies in cadastre building definitions. For example, some buildings that represent modern apartment blocks are split into multiple touching polygons representing the physical parts of the building such as an entrance; or an administrative part such as an address. Other modern apartment blocks, often in the same city, are presented as a single polygon even though they may have the same number of physical units (entrances) or administrative addresses associated with them.

To calculate it, the facade ratio - area over perimeter - is measured for each group of adjacent buildings within the morphotope. The final value of the "Likely Occupied Area" is 1 if more than 40% of the built up area with no courtyards has a facade ratio of greater than four and elongation less than .9 and furthermore.

The "Area of the largest ten connected structures" is calculated by summing the area of the largest ten connected structures in the morphotope. It is a measure of the built up area within the morphotope and is used for the final clustering.

Similarly, The "Perimeter of the largest ten connected structures" is calculated by summing the perimeters of the largest ten connected structures in the morphotope. It is not directly used in the taxonomy construction - it function is to detect outliers based on faulty cadastre data.

## B. Spatial Agglomerative Adaptive Aggregation (SA3)

Spatial Agglomerative Adaptive Aggregation (SA3) relies combines two disparate clustering approaches.

1. First, it generates a full Ward clustering tree based on differences in feature space, and adjacency in geographic space.
2. Second it uses Leaf extraction to generate a set of clusters from the dendrogram.

The linkage matrix is generated by computing distances between observations based on the Ward formula, subject to a restriction that new connections must be spatially adjacent enclosed tessellation cells. The leaf extraction algorithm processes the resulting dendrogram as follows:

1. The dendrogram is cut at all possible levels - one for each connection - starting from the lowest to the highest distance value.
2. At every level the number of members within each cluster and its constituent children are counted. If a cluster has more than N ETCs it is marked for extraction.
3. When one cluster marked for extraction merges with another, both are extracted from the dendrogram as separate clusters. Since the members of a marked cluster keeps increasing until a merger occurs, typically each extracted cluster has more than N members.
4. All points that are never part of a marked cluster are treated as outliers and marked as noise.

Before applying the clustering algorithm, the all variables are preprocessed using a Quantile Transformer with a uniform distribution. This data transformation produces a relatively more equal weighing of all features when calculating distances between observations, as opposed to standardisation or normalisation. For example, variables with large ranges such as area, dominate less the distance calculations in feature space.

## C. Data Normalisation & Outlier exclusion

Before the taxonomic tree is constructed, the data features have to be normalised to ensure individual variables do not dominate the distance calculations. We therefore drop any outliers that might affect this transformation. The need to do this arises again from the heterogenous data quality in the cadastres of different countries. In some cases entire industrial areas are delineated as single buildings, in others power lines or sheds are present in the cadastre.

Specifically, we re-assign as noise any morphotopes that have either - a "Perimeter of the largest ten connected structures" larger than 200 km.; or "Area of the largest ten connected structures" larger than 500 sq.m.; or median building area less than 20 sq.m; or median building perimeter larger than 5km.

Once the outliers are re-assigned to noise, the rest of the data features are standardised. The outliers are processed together with the rest of the noise ETCs in subsequent steps.

## D. Data preprocessing

### D.1. Building preprocessing pipeline

- The first step in the is to split up multi-polygons and make the geometries valid.
- The second step, is to simplify the polygons in order to accurately represent the corners of buildings and other shape related characteristics.
- Next, to filter out any buildings that have an area larger than 200,000 sq.m. This is done since some artefacts such as construction sites, mines or tunnels might be included in the data as buildings.
- The next step is to merge overlapping buildings that either: overlap for at least 10 percent of their areas, or one of them has less than 50 sq.m. in total area. This is done to merge buildings and building parts, since cadastre definitions of these two polygon types are inconsistent and sometimes buildings are assigned as building parts or vice versa. This step merges the buildings and its parts into one polygon.
- Finally, the preprocessing pipeline snaps nearby buildings together and fills gaps in the polygons that are less than 10 square cm. These two steps aim to address some common topological issues, such as missing slivers with almost zero areas between multiple or inside individual building polygons. Nevertheless, even after the preprocessing numerous topological issues remain and therefore we take this into account in subsequent analysis steps.

### D.2. Street segments preprocessing pipeline

| *The types of street segments included in the analysis* |
| --- |
| living_street |
| motorway |
| motorway_link |
| pedestrian |
| primary |
| primary_link |
| residential |
| secondary |
| secondary_link |
| tertiary |
| tertiary_link |
| trunk |
| trunk_link |
| unclassified |

The second major stage of the street processing is the simplification of the street network. The data from OvertureMaps, focuses on representation of street network for transportation purposes. That means it tends to include multiple geometries for wide boulevards where each captures a single carriageway, complex representation of junctions or even the smallest artefacts of transportation-based focus. However, such a network is not directly usable for morphological analysis as it does not capture morphological perception of street network which is usually captured via street centrelines, omitting transportation detail. For this reason, we apply the simplification method based on the detection of the problematic parts of the network (Fleischmann and Vybornova 2024). This ensures automatised algorithmic cleaning of street networks resulting in a morphological representation derived from the transportation one - dual carriage ways are merged together, roundabouts are simplified, and other changes.

### D.3. Cadastral Buildings

Even though the data for each country comes from their respective official cadastre, there are still issues present.

First, there are cases where official cadastre data on building polygons within the same country comes from various sources, is in different formats and has inconsistent definitions of what a building is and how it should be split into units. Therefore, the data for each country or region is downloaded individually and converted to the same format and only two-dimensional building polygons and no other properties are used for the analysis.

The inconsistent delineation issue is partially addressed through the design and selection of morphometric attributes. An extra step, specific only to Czechia is carried out to further address this. The boundaries of modernist housing estates in the official cadastre data are dictated by the coverage of local housing associations and not physical features. As such there is no consistency in their delineation across the country. To remedy this to some extent, we use a dataset on the spatial positions of communist era large scale developments and merge together the polygons that fall within them for consistency.

As a last step, we approximate building adjacency by assuming that any building within 50cm of another building is touching it and carry out all character calculations based on this assumption. This is done to avoid other subtle topological and numerical issues when dealing with high precision data.

### D.4. Overture Maps street network data

The street network is a direct download from Overture Maps Transportation theme, a processed subset of data from OpenStreetMap, which has global coverage and high quality data. It is processed in two steps.

First, since the dataset includes multiple segment types, including footpaths and service roads that represent parking lot spaces, we limit the analysis to the segment types specified above. Another type of segment that is filtered out are tunnels - the analysis strictly focuses on two dimensions and therefore undergrounds structures adversely affect the calculation of boundaries and characters.

The second major stage of the street processing is the simplification of the street network. The data from OvertureMaps, focuses on representation of street network for transportation purposes. That means it tends to include multiple geometries for wide boulevards where each captures a single carriageway, complex representation of junctions or even the smallest artefacts of transportation-based focus. This information is secondary for morphological analysis and can adversely affect the calculated street attributes, therefore the network is standardised.

## E. Cluster pen portraits

### E.1. Level 1

| Cluster name | Cluster Description |
|---|---|
| **Incoherent Fabric** | Incoherent fabric covers a wide morphological variety, with a common theme of partial or complete breakage of the traditional structural roles of streets, plots, and buildings. Common for modernist period, post-modern, and industrial developments, this branch typically has a less connected street network and may showcase buildings facing open spaces and internal parts |

| Cluster name | Cluster Description |
| --- | --- |
|  | of blocks rather than streets. At the same time, it contains less defined village developments. |
| **Coherent Fabric** | In coherent fabric, all streets, plots, and buildings take their traditional structural roles in defining the spatial arrangement of the urban form. It is common for traditional European development with densely connected street networks and legible plot structure, facilitating direct relation between buildings and streets. |

## E.2. Level 2

| Cluster name | Cluster Description |
| --- | --- |
| **Incoherent Large-scale Fabric** | Incoherent large-scale fabric captures typically urban development composed of buildings larger than the average, that may or may not be far from each other, creating large open spaces. Streets tend to be of an utilitarian use, rather than a structural one, typical for modernist housing estates or industrial zones. |
| **Incoherent Small-scale Fabric** | Incoherent small-scale fabric is mostly non-urban development capturing various kinds of villages and small towns, which show high variation of morphological properties. Buildings tend to be smaller, but distances between them vary, as well as the relations between buildings and streets. |
| **Coherent Interconnected Fabric** | This cluster has very high built-up density and local street connectivity, with narrow and short streets. It is primarily characterised by a high count of connected buildings forming enclosed blocks with courtyards. Furthermore, the distances between the formed blocks is small. |
| **Coherent Dense Fabric** | Dense fabric captures morphology typical for urban residential areas with lower density, where blocks are defined by streets more than buildings. The street networks are well defined and connected with buildings being either adjacent (e.g. row houses) or disjoint (e.g. urban villas). |

## E.3. Level 3

| Cluster name | Cluster Description |
| --- | --- |
| **Incoherent Large-scale Homogeneous Fabric** | This cluster consists of the large buildings with moderate variations in size and shape, as well as low to moderate street connectivity and wide streets. The resulting environment is spacious, with significant open areas between structures, typical of modernist housing. |
| **Incoherent Large-scale Heterogeneous Fabric** | This cluster consists of the largest buildings with notable variations in size and shape, as well as low to moderate street connectivity and wide streets. The design does not emphasize sunlight exposure, creating broad but less refined configurations, typical of industrial and other service areas. |
| **Incoherent Small-scale Linear Fabric** | This cluster has a moderate built up area and the low local street connectivity in the taxonomy, typically forming long linear villages. Its streets are long, linear, wide and there are minimal shared walls between structures. |
| **Incoherent Small-scale Sparse Fabric** | This cluster is characterized by low built-up density, low street connectivity, large distances between buildings, few shared walls, and large open spaces around buildings. The streets are few, open and wide. The |

| Cluster name | Cluster Description |
|---|---|
| | buildings are small to moderate in size and their layout is more typical of rural areas. |
| **Incoherent Small-scale Compact fabric** | This cluster has low to moderate built-up area and street connectivity. Buildings exhibit a consistent alignment among themselves and also along streets of varying length, width and linearity. There is also a significant number of shared walls between structures. |
| **Coherent Interconnected Fabric** | This cluster has very high built-up density and local street connectivity, with narrow and short streets. It is primarily characterised by a high count of connected buildings forming enclosed blocks with courtyards. Furthermore, the distances between the formed blocks is small. |
| **Coherent Dense Disjoint Fabric** | This cluster has moderate to high built-up density and local street connectivity, with longer and wider streets compared to other dense developments. Shared walls between buildings are less common, and distances within buildings are moderate, reflecting a pattern of stand-alone structures within a robust street network . |
| **Coherent Dense Adjacent Fabric** | In this cluster, the built-up density and local street connectivity are high, while inter-building distances remain relatively small. Buildings frequently share walls, forming larger structures along relatively short and narrow streets. |

### E.4. Feature matrices of clusters

## F. Adaptive limit for tessellation

Enclosed tessellation typically covers entirety of study area. In our case, we need to limit it only to those cells belonging to buildings while minimising the boundary effect on the edges of settlements generating unnaturally large cells. Hence, the delineation is further modified by introducing a variable, individual bandwidth for every building. The limits used here are calculated through a Gabriel graph-based filtering of the Voronoi graph of building centroids, which takes into account the surrounding neighbours structure around every building. This process results in a local bandwidth for every building based on distances to neighbouring buildings computed as the half of the maximum distance between neighbours of each building + 10% of the maximum distance. For example, in row housing the bandwidth will be relatively small, in single family housing estates the bandwidth will be larger, and in industrial areas larger even more; regardless of whether or not these buildings are in the middle of cities or around their edges.

## G. Comparison

### G.1. CORINE Land Cover classification

Figure A1 shows the distribution of the HiMoC level three clusters across CORINE Land Cover classes. In general, the branches of built-up fabric broadly follow the higher level patterns that are described by the CORINE Land Cover Classification. However, our branches split the CORINE classes into different sub-classes, providing more information on urban form.

| | Continuous urban fabric | Discontinuous urban fabric | Industrial or commercial units | Non-irrigated arable land | Pastures | Complex cultivation patterns | Land principally occupied by agriculture, with significant areas of natural vegetation |
|---|---|---|---|---|---|---|---|
| Incoherent Large-Scale Homogeneous Fabric | 0.03 | 0.61 | 0.23 | 0.05 | 0.02 | 0.01 | 0.01 |
| Incoherent Large-Scale Heterogeneous Fabric | 0.01 | 0.30 | 0.36 | 0.16 | 0.10 | 0.01 | 0.01 |
| Incoherent Small-Scale Linear Fabric | 0.00 | 0.29 | 0.01 | 0.28 | 0.13 | 0.13 | 0.06 |
| Incoherent Small-Scale Sparse Fabric | 0.00 | 0.36 | 0.01 | 0.21 | 0.14 | 0.13 | 0.07 |
| Incoherent Small-Scale Compact Fabric | 0.00 | 0.68 | 0.02 | 0.08 | 0.11 | 0.03 | 0.02 |
| Coherent Interconnected Fabric | 0.46 | 0.49 | 0.04 | 0.00 | 0.00 | 0.00 | 0.00 |
| Coherent Dense Disjoint Fabric | 0.01 | 0.88 | 0.01 | 0.03 | 0.04 | 0.00 | 0.00 |
| Coherent Dense Adjacent Fabric | 0.07 | 0.85 | 0.03 | 0.02 | 0.02 | 0.00 | 0.00 |

*Figure A1.* CORINE Land Cover cross-tabulation showing the proportion of each branch of built-up fabric within relevant CORINE classes.

"Discontinuous Urban Fabric" is split into multiple classes based on the density, building adjacency, street characteristics and overall distances between buildings by the clusters – *Incoherent Large-scale Homogeneous Fabric*, *Incoherent Small-Scale Compact Fabric*, *Coherent Dense Disjoint Fabric* and *Coherent Dense Adjacent Fabric*. It is worth noting that both *Coherent Dense Disjoint Fabric* and *Coherent Dense Adjacent Fabric* are mostly discontinuous urban land, with *Coherent Interconnected Fabric* being nearly the only branch present within "Continuous Urban Fabric" class of CORINE. However, only half of it is "Continuous Urban Fabric", indicating that the two concepts are not perfectly aligned.

The *Incoherent Small-Scale Sparse Fabric* and *Incoherent Small-Scale Linear Fabric* classes have the least amount of urban land and overlap with some agriculture and developed land classes. This is likely due to their presence within many rural areas. The *Incoherent Small-Scale Compact Fabric* stands in between these two more rural classes and the more urban ones. *Incoherent Large-scale Heterogeneous Fabric* overlaps the most with the CORINE's Industrial and Commercial Units class, which is expected and highlights that heterogeneity at large scale in urban environments is typically serving non-residential purposes.

### G.2. Copernicus Urban Atlas

Figure A2 shows a similar comparison with the Copernicus Urban Atlas. The results follow very similar patterns to the CORINE Land Cover comparisons, with some interesting differences due to the greater granularity of the Urban Atlas inside urban areas.

| | Continuous urban fabric (S.L. : > 80%) | Discontinuous dense urban fabric (S.L. : 50% - 80%) | Discontinuous low density urban fabric (S.L. : 10% - 30%) | Discontinuous medium density urban fabric (S.L. : 30% - 50%) | Discontinuous very low density urban fabric (S.L. : < 10%) | Industrial, commercial, public, military and private units | Isolated structures |
|---|---|---|---|---|---|---|---|
| **Incoherent Large-Scale Homogeneous Fabric** | 0.19 | 0.26 | 0.01 | 0.06 | 0.00 | 0.39 | 0.01 |
| **Incoherent Large-Scale Heterogeneous Fabric** | 0.05 | 0.14 | 0.03 | 0.07 | 0.02 | 0.51 | 0.05 |
| **Incoherent Small-Scale Linear Fabric** | 0.04 | 0.22 | 0.09 | 0.13 | 0.05 | 0.11 | 0.20 |
| **Incoherent Small-Scale Sparse Fabric** | 0.04 | 0.23 | 0.10 | 0.15 | 0.07 | 0.10 | 0.14 |
| **Incoherent Small-Scale Compact Fabric** | 0.06 | 0.29 | 0.11 | 0.22 | 0.03 | 0.07 | 0.02 |
| **Coherent Interconnected Fabric** | 0.69 | 0.18 | 0.00 | 0.02 | 0.00 | 0.09 | 0.00 |
| **Coherent Dense Disjoint Fabric** | 0.14 | 0.51 | 0.05 | 0.20 | 0.01 | 0.04 | 0.00 |
| **Coherent Dense Adjacent Fabric** | 0.34 | 0.46 | 0.01 | 0.08 | 0.00 | 0.07 | 0.00 |

*Figure A2.* Copernicus Urban Atlas cross-tabulation showing the proportion of each branch of built-up fabric within relevant Urban Atlas classes.

Similarly to the CORINE comparison, *Incoherent Large-scale Heterogeneous Fabric* has a large overlap with industrial buildings. *Incoherent Large-scale Homogeneous Fabric* also intersects the most with this class, but 60% is split across the rest of the urban classes as these are often residential structures of typically modernist spatial organisation, but not exclusively. *Incoherent Small-Scale Sparse Fabric* and *Incoherent Small-Scale Linear Fabric* overlap with the "Isolated structures" and lower density classes but none shows a dominance, reflecting a relatively weak alignment of their definition to the continuity-density definition used in Urban Atlas. *Incoherent Small-scale Compact Fabric* again stands in between these two more sparse classes and the more urban ones. In contrast to the CORINE Land Cover comparison, the *Coherent Interconnected Fabric* cluster is mainly falling in "Continuous Urban Fabric", reflecting the fact that it covers mainly many interconnected buildings. *Coherent Dense Adjacent Fabric* is split between the Continuous and Discontinuous dense classes, in contrast to the previous comparison where it was mainly discontinuous. This indicates that not only are there differences between our classes and existing data products but also a degree of disagreement between the data products themselves.

### G.3. Local Climate Zones table

Finally, Figure A3 illustrates the comparison with Local Climate Zones using the prediction by Demuzere et al. (2022). This broadly follows the same patterns as above. However, to put the comparison in context, around 47% of the Central European buildings are assigned to the "Open low-rise" LCZ class, with the second most common class "Sparsely built" having around 27% of the rest and the third one, "Low plants", with 10%. Our most common classes are *Incoherent Small-Scale Compact Fabric* with 42%, *Coherent Dense Disjoint Fabric* and *Incoherent Small-Scale Sparse Fabric* with 15% each. Another thing to note is that there are a lot of buildings and fabrics that are classified in LCZ as part of the non-developed class, such as variations of plans and trees, which indicates the inability of LCZ to properly capture typically smaller settlements.

| | Compact high-rise | Compact mid-rise | Compact low-rise | Open high-rise | Open mid-rise | Open low-rise | Lightweight low-rise | Large low-rise | Sparsely built | Heavy industry | Dense trees | Scattered trees | Bush, scrub | Low plants | Bare rock or paved | Bare soil or sand | Water |
|---|---|---|---|---|---|---|---|---|---|---|---|---|---|---|---|---|---|
| **Incoherent Large-Scale Homogeneous Fabric** | 0.00 | 0.01 | 0.00 | 0.01 | 0.15 | 0.55 | 0.00 | 0.11 | 0.09 | 0.00 | 0.01 | 0.02 | 0.00 | 0.05 | 0.00 | 0.00 | 0.00 |
| **Incoherent Large-Scale Heterogeneous Fabric** | 0.00 | 0.00 | 0.00 | 0.00 | 0.04 | 0.28 | 0.00 | 0.24 | 0.19 | 0.00 | 0.03 | 0.05 | 0.00 | 0.14 | 0.00 | 0.00 | 0.00 |
| **Incoherent Small-Scale Linear Fabric** | 0.00 | 0.00 | 0.00 | 0.00 | 0.00 | 0.04 | 0.00 | 0.00 | 0.37 | 0.00 | 0.07 | 0.14 | 0.00 | 0.38 | 0.00 | 0.00 | 0.00 |
| **Incoherent Small-Scale Sparse Fabric** | 0.00 | 0.00 | 0.00 | 0.00 | 0.00 | 0.10 | 0.00 | 0.00 | 0.41 | 0.00 | 0.07 | 0.17 | 0.00 | 0.24 | 0.00 | 0.00 | 0.00 |
| **Incoherent Small-Scale Compact Fabric** | 0.00 | 0.00 | 0.00 | 0.00 | 0.01 | 0.34 | 0.00 | 0.01 | 0.44 | 0.00 | 0.04 | 0.07 | 0.00 | 0.09 | 0.00 | 0.00 | 0.00 |
| **Coherent Interconnected Fabric** | 0.01 | 0.22 | 0.03 | 0.03 | 0.31 | 0.34 | 0.00 | 0.04 | 0.02 | 0.00 | 0.00 | 0.00 | 0.00 | 0.00 | 0.00 | 0.00 | 0.00 |
| **Coherent Dense Disjoint Fabric** | 0.00 | 0.00 | 0.00 | 0.00 | 0.02 | 0.68 | 0.00 | 0.01 | 0.23 | 0.00 | 0.02 | 0.02 | 0.00 | 0.03 | 0.00 | 0.00 | 0.00 |
| **Coherent Dense Adjacent Fabric** | 0.00 | 0.01 | 0.01 | 0.00 | 0.08 | 0.78 | 0.00 | 0.02 | 0.08 | 0.00 | 0.01 | 0.00 | 0.00 | 0.01 | 0.00 | 0.00 | 0.00 |

*Figure A3.* Local Climate Zones cross-tabulation showing the proportion of each branch of built-up fabric within relevant LCZ.

Similarly to the other comparisons, *Incoherent Large-scale Heterogeneous Fabric* has a large overlap with large low-rise buildings, representing the industrial areas. *Coherent Dense Disjoint Fabric*, *Coherent Dense Adjacent Fabric* and *Incoherent Large-scale Homogeneous Fabric* show the highest overlap with "Open low-rise", suggesting that we are splitting this cluster based on characteristics that LCZ does not consider. *Coherent Interconnected Fabric* is the only cluster that has any overlap with compact mid-rise, but still is majorly open low-rise. *Incoherent Small-scale Sparse Fabric* and *Incoherent Small-scale Linear Fabric* overlap mostly with the "Sparsely built" cluster, as both cover primarily rural areas.